\documentclass[letterpaper, 10 pt, conference]{ieeeconf}  

\IEEEoverridecommandlockouts                              

\usepackage{graphicx} 
\usepackage{amsmath} 
\usepackage{amssymb}  
\usepackage{bm}
\usepackage{color}
\usepackage[font=footnotesize]{subcaption}
\usepackage{comment}
\usepackage[font=footnotesize]{caption}
\usepackage{siunitx}
\usepackage[linesnumbered,ruled]{algorithm2e}
\newcommand{\RomanNumeralCaps}[1] 
    {\MakeUppercase{\romannumeral #1}} 
\newtheorem{theorem}{Theorem}

\newtheorem{remark}{Remark}
\DeclareMathOperator*{\argmin}{arg\,min}

\usepackage[style=ieee]{biblatex}
\renewbibmacro*{issue+date}{%
  \setunit{\addcomma\space}
    \iffieldundef{issue}
      {\usebibmacro{date}}
      {\printfield{issue}%
      \setunit*{\addspace}%
      \usebibmacro{date}}
  \newunit}

\newcommand{\M}[1]{\ensuremath{\textbf{\text{#1}}}} 
\newcommand{\V}[1]{\ensuremath{\textbf{\text{#1}}}} 
\newcommand{\Exp}[2]{\ensuremath{\mathop{\mathbb{E}}_{\substack{{#1}}} \left[{#2}\right]}} 
\newcommand{\Var}{\ensuremath{\operatorname{Var}}} 
\newcommand{\tr}[1]{{#1}^{\ensuremath{\mathsf{T}}}} 
\newcommand{\inv}[1]{{#1}^{\ensuremath{\!\!\mathsf{-1}\,}}} 
\newcommand{\Th}{\ensuremath{{}^{\textrm{th}}}}

\title{\LARGE \bf
Decoupling stochastic optimal control problems for efficient solution: insights from experiments across a wide range of noise regimes
}

\author{Mohamed Naveed Gul Mohamed$^{1}$, Suman Chakravorty$^{1}$ and Dylan A. Shell$^{2}$
\thanks{$^{1}$M. N. Gul Mohamed and S. Chakravorty are with the Department of Aerospace Engineering, and $^{2}$D. A. Shell is with the Department of Computer Science \& Engineering, Texas A\&M University, College Station, TX 77843 USA. \{\tt mohdnaveed96@gmail.com, schakrav@tamu.edu, dshell@tamu.edu\}}
}

\begin{document}

\maketitle
\thispagestyle{empty}
\pagestyle{empty}

\begin{abstract}
We consider the problem of robotic planning under uncertainty in this paper. This problem may be posed as a stochastic optimal control problem, a solution to which is fundamentally intractable owing to the infamous ``curse of dimensionality''. Hence, we consider the extension of a ``decoupling principle'' that was recently proposed by some of the authors, wherein a nominal open-loop problem is solved followed by a linear feedback design around the open-loop, and which was shown to be near-optimal to second order in terms of a ``small noise" parameter, to a much wider range of noise levels. Our empirical evidence suggests that this allows for tractable planning over a wide range of uncertainty conditions without unduly sacrificing performance.
\end{abstract}

\section{Introduction}

Planning under uncertainty is a central problem in robotics.  The space of
current methods includes several contenders, each with different simplifying
assumptions, approximations, and domains of applicability.  This is a natural
consequence of the fact that the challenge of dealing with the continuous
state, control and observation space problems, for non-linear systems and
across long-time horizons with significant noise, and potentially multiple
agents, is fundamentally intractable.

%
 
\begin{figure}
    \centering
    \begin{subfigure}[b]{0.225\textwidth}
        \centering
        \includegraphics[width=\textwidth]{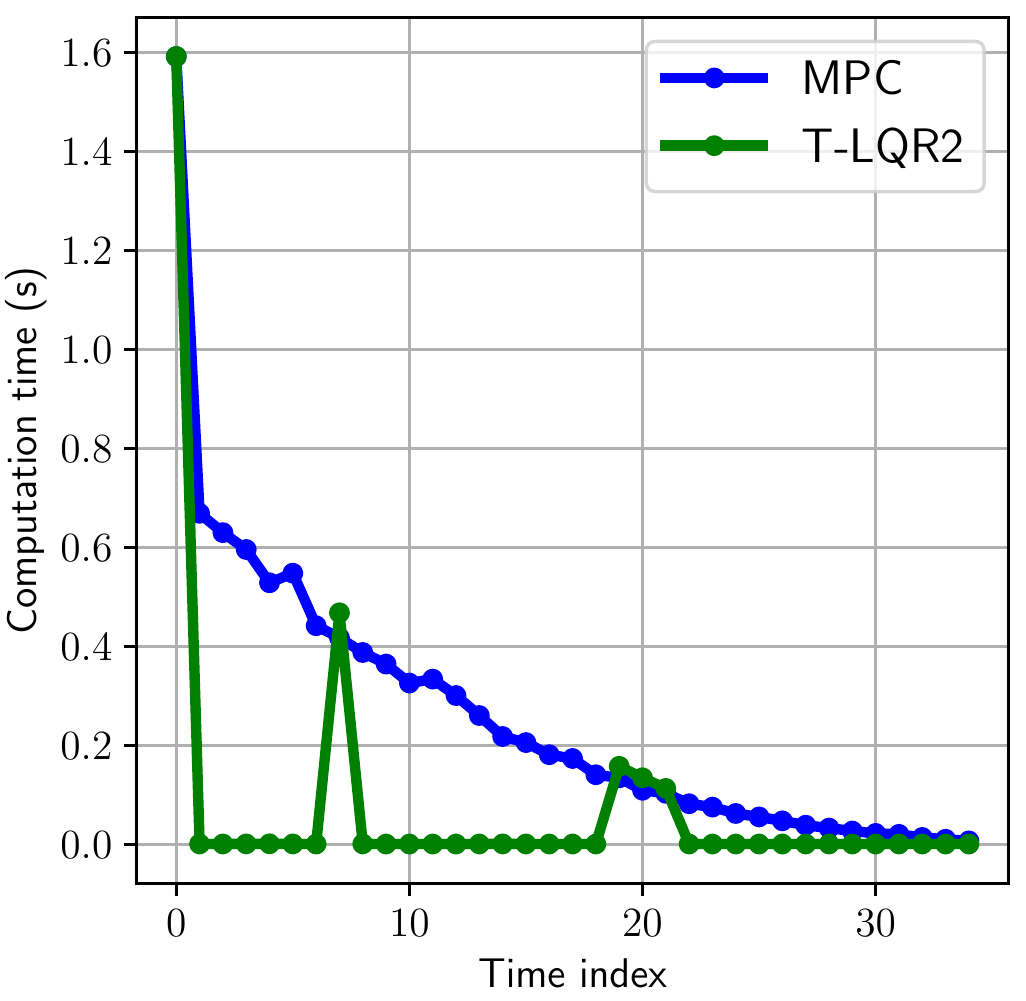}
        \caption{A single agent.}
        \label{1_agent_replan_time}
    \end{subfigure}
    \begin{subfigure}[b]{0.225\textwidth}
        \centering
        \includegraphics[width=\textwidth]{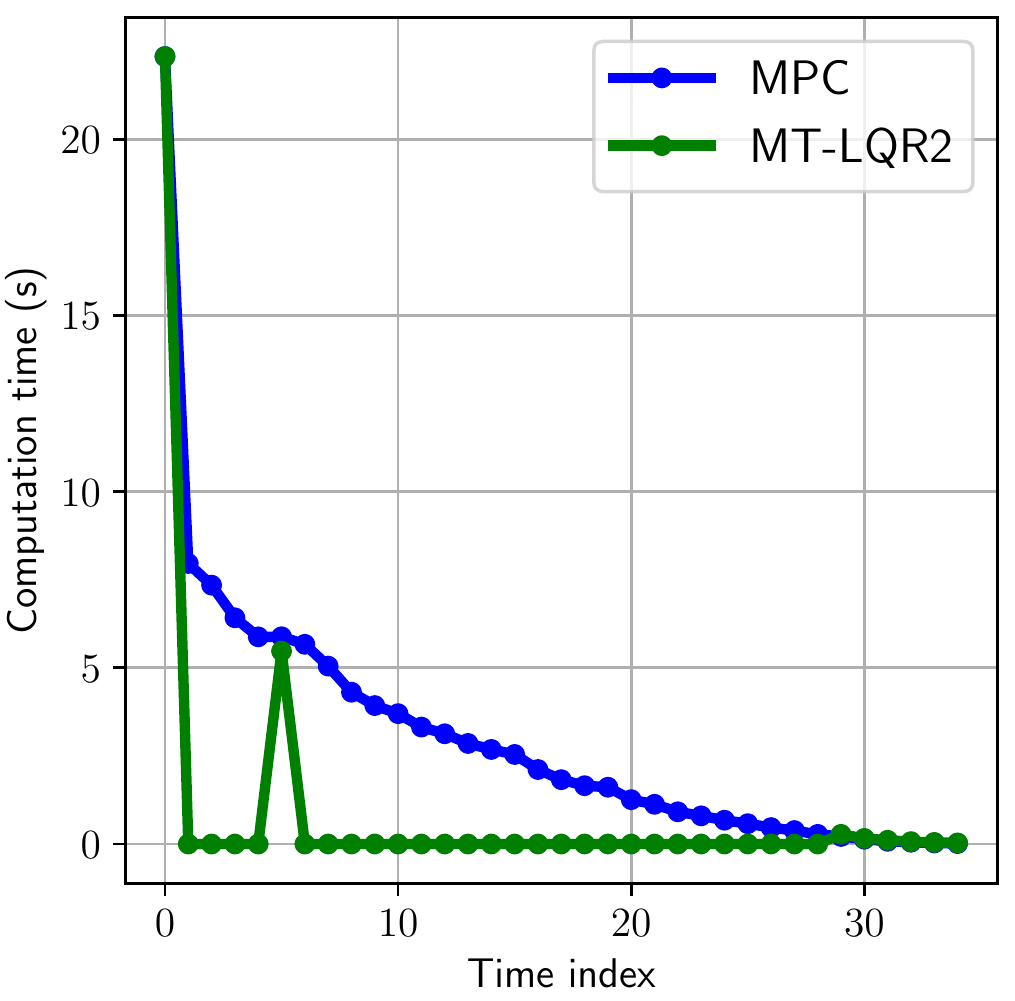}
        \caption{Three agents.}
        \label{3_agent_replan_time}
    \end{subfigure}
    \caption{Computation time expended by MPC (in blue) and the algorithms we describe (in green), at each time step for a sample experiment involving navigation. Both cases result in nearly identical motions by the robot.
        The peaks in T-LQR2 and MT-LQR2 happen only when replanning takes place. Computational effort decreases for both methods because the horizon diminishes as the agent(s) reach their goals.
    (To relate to subsequent figures: noise parameter $\epsilon = 0.4$ and the replan threshold = 2\% of cost deviation.)}
    \label{fig:timeplot}
\end{figure}

Model Predictive Control is one popular means for tackling optimal control
problems~\cite{Mayne_1,Mayne_2}.  The MPC approach solves a finite horizon
``deterministic'' optimal control problem at every time step given the current
state of the process, performs only the first control action and then repeats
the planning process at the next time step. In terms of computation, this is a
costly endeavor.  When a stochastic control problem is well approximated by the
deterministic problem, namely when the noise is meager, much of this
computation is simply superfluous.  In this paper we consider a recently
proposed method~\cite{D2C1.0}, grounded on a decoupling result, that uses a local
feedback to control noise induced deviations from the deterministic (that we term
the ``nominal'') trajectory. When the deviation is too large for the feedback
to manage, replanning is triggered and it computes a fresh nominal.  Otherwise,
the feedback tames the perturbations during execution and no computation is
expended in replanning.  Figure~\ref{fig:timeplot} illustrates this: the
areas under the respective curves give the total computational resources
consumed---the savings are seen to be considerable.

This paper presents an empirical investigation of this decoupling approach,
exploring dimensions that are important in characterizing its performance.
The primary focus is on understanding the performance across a wide range of 
noise conditions.

%




\subsection{Related Work}

Robotic planning problems under uncertainty can be posed as a stochastic optimal control problem that requires the solution of an associated Dynamic Programming (DP) problem, however, as the state dimension $d$ increases, the computational complexity
goes up exponentially \cite{bertsekas1}, Bellman's infamous ``curse of dimensionality".  There has been recent success using sophisticated (Deep) Reinforcement Learning (RL) paradigm to solve DP problems, where deep neural networks are used as the function approximators \cite{RLHD1, RLHD2,RLHD3, RLHD4, RLHD5}, however, the training time required for these approaches is still prohibitive to permit real-time robotic planning that is considered here. 
In the case of continuous state, control and observation space problems, the
Model Predictive Control \cite{Mayne_1, Mayne_2} approach has been used with a
lot of success in the control system and robotics community.   For deterministic
systems, the process results in solving the original DP problem in a recursive
online fashion. However, stochastic control problems, and the control of
uncertain systems in general, is still an unresolved problem in MPC. As
succinctly noted in \cite{Mayne_1}, the problem arises due to
the fact that in stochastic control problems, the MPC optimization at every
time step cannot be over deterministic control sequences, but rather has to be
over feedback policies, which is, in general, difficult to
accomplish because a compact, tractable parametrization of such policies to perform the optimization is, in general, unavailable. Thus, the tube-based MPC approach, and its stochastic counterparts,
typically consider linear systems \cite{T-MPC1, T-MPC2,T-MPC3} for which a
linear parametrization of the feedback policy suffices but the methods require
expensive offline computation when dealing with nonlinear systems. In recent work, we have introduced a ``decoupling principle" that allows us to tractably solve such stochastic optimal control problems in a near optimal fashion, with applications to highly efficient RL and MPC implementations \cite{D2C1.0,T-PFC}. However, this prior work required a small noise assumption. In this work, we relax this small noise assumption to show, via extensive empirical evaluation, that even when the noise is not small, a ``replanning" modification of the decoupled planning algorithms suffice to keep the planning computationally efficient while retaining performance comparable to MPC. 

The problem of multiple agents further and severely compounds the planning
problem since now we are also faced with the issue of a control space that
grows exponentially with the number of agents in the system. Moreover, since
the individual agents never have full information regarding the system state,
the observations are partial. Furthermore, the decision making has to be done
in a distributed fashion which places additional constraints on the networking
and communication resources.  In a multi-agent setting, the stochastic optimal
problem can be formulated in the space of joint policies.  Some variations of
this problem have been successfully characterized and tackled based on the
level of observability, in/dependence of the dynamics, cost functions and
communications
\cite{seuken2008formal,oliehoek2016concise,pynadath2002communicative}. This has
resulted in a variety of solutions from fully-centralized
\cite{boutilier1996planning} to fully-decentralized approaches with many
different subclasses \cite{amato2013decentralizedB,oliehoek2012decentralized}.
The major concerns of the
multi-agent problem are tractability of the solution and the level of
communication required during the execution of the policies. In this paper, we shall consider a generalization of the decoupling principle to a multi-agent, fully observed setting. We show that this leads to a spatial decoupling between agents in that they do not need to communicate for long periods of time during execution. Albeit, we do not consider the problem of when and how to replan in this paper, assuming that there exists a (yet to be determined) distributed mechanism that can achieve this, we nonetheless show that there is a highly significant increase in planning efficiency over a wide range of noise levels.

\subsection{Outline of Paper}
The rest of the document is organised as follows: Section~\ref{section:prob} states the problem, \ref{section:decoupling} gives background on the decoupling principle, \RomanNumeralCaps{4} explains the planning algorithms used, \RomanNumeralCaps{5} discusses the results and observations and  \ref{section:conclusion} concludes.  
\section{Problem Formulation}
\label{section:prob}
The problem of robot planning and control under noise can be formulated as a stochastic optimal control problem in the space of feedback policies. We assume here that the map of the environment is known and state of the robot is fully observed. Uncertainty in the problem lies in the system's actions. %
\subsection{System Model:}
For a dynamic system, we denote the state and control vectors by $\V{x}_t \in \ \mathbb{X} \subset \ \mathbb{R}^{n_x}$ and $\V{u}_t \in \ \mathbb{U} \subset \ \mathbb{R}^{n_u}$ respectively at time $t$. The motion model $f : \mathbb{X} \times \mathbb{U} \times \mathbb{R}^{n_u}   \rightarrow \mathbb{X} $ is given by the equation 
\begin{equation}
    \V{x}_{t+1}= f(\V{x}_t, \V{u}_t, \epsilon\V{w}_t); \  \V{w}_t \sim \mathcal{N}(\V{0}, {\mathbf \Sigma}_{\V{w}_t}) 
    \label{eq:model},
\end{equation}
where \{$\V{w}_t$\} are zero mean independent, identically distributed (i.i.d) random sequences with variance ${\mathbf\Sigma}_{\V{w}_t}$, and $\epsilon$ is a small parameter modulating the noise input to the system. 
\subsection{Stochastic optimal control problem:} %
The stochastic optimal control problem for a dynamic system with initial state $\V{x}_0$ is defined as:
\begin{equation}
    J_{\pi^{*}}(\V{x}_0) = \min_{\pi} \ \Exp{}{\sum^{T-1}_{t=0} c(\V{x}_t, \pi_t (\V{x}_t)) + c_T(\V{x}_T)},
\end{equation}
\begin{equation}
   s.t.\ \V{x}_{t+1} = f(\V{x}_t, \pi_t (\V{x}_t), \epsilon\V{w}_t),
\end{equation}
where:
\begin{itemize}
    \item the optimization is over feedback policies $\pi := \{ \pi_0, \pi_1, \ldots, \pi_{T-1} \} $ and $\pi_t(\cdot)$: $\mathbb{X} \rightarrow \mathbb{U}$ specifies an action given the state, $\V{u}_t = \pi_t(\V{x}_t)$;
    \item $J_{\pi^{*}}(\cdot): \mathbb{X} \rightarrow \mathbb{R}$  is the cost function when the optimal policy $\pi^{*}$ is executed; 
    \item $c_t(\cdot,\cdot): \mathbb{X} \times \mathbb{U} \rightarrow \mathbb{R} $  is the one-step cost function;
    \item $c_T(\cdot): \mathbb{X} \rightarrow \mathbb{R}$ is the terminal cost function;
    \item $T$ is the horizon of the problem;
    \item the expectation is taken over the random variable $\V{w}_t$.
\end{itemize}
\section{A Decoupling Principle}
\label{section:decoupling}
Now, we give a brief overview of a ``decoupling principle'' that allows us to substantially reduce the complexity of the stochastic planning problem given that the parameter $\epsilon$ is small enough. We only provide an outline here and the relevant details can be found in our recent work \cite{D2C1.0}. We shall also present a generalization to a class of multi-robot problems. Finally, we preview the results in the rest of the paper.
\subsection{Near-Optimal Decoupling in Stochastic Optimal Control}
Let $\pi_t(\V{x}_t)$ denote a control policy for the stochastic planning problem above, not necessarily the optimal policy. Consider now the control actions of the policy when the noise to the system is uniformly zero, and let us denote the resulting ``nominal'' trajectory and controls as $\overline{\V{x}}_t$ and $\overline{\V{u}}_t$ respectively, i.e., $\overline{\V{x}}_{t+1} = f(\overline{\V{x}}_t, \overline{\V{u}}_t, 0)$, where $\overline{\V{u}}_t = \pi_t(\overline{\V{x}}_t)$. Note that this nominal system is well defined. \\
Further, let us assume that the closed-loop (i.e., with $\V{u}_t = \pi_t(\V{x}_t)$), system equations, and the feedback law are smooth enough that we can expand the feedback law about the nominal as $\pi_t(\V{x}_t) = \overline{\V{u}}_t + \M{K}_t\delta \V{x}_t + \M{R}_t^{\pi}(\delta \V{x}_t)$, where $\delta \V{x}_t = \V{x}_t - \overline{\V{x}}_t$, i.e., the perturbation from the nominal, $\M{K}_t$ is the linear gain obtained by the Taylor expansion about the nominal in terms of the perturbation $\delta \V{x}_t$, and $\M{R}_t^{\pi}(\cdot)$ represents the second and higher order terms in the expansion of the feedback law about the nominal trajectory. Further we assume that the closed-loop perturbation state can be expanded about the nominal as: $\delta \V{x}_t = \M{A}_t \delta \V{x}_t + \M{B}_t \M{K}_t \delta \V{x}_t + \M{R}_t^f (\delta \V{x}_t) + \epsilon \M{B}_t \V{w}_t$, where the $\M{A}_t$, $\M{B}_t$ are the system matrices obtained by linearizing the system state equations about the nominal state and control, while $\M{R}_t^f(\cdot)$ represents the second and higher order terms in the closed-loop dynamics in terms of the state perturbation $\delta \V{x}_t$. Moreover, let the nominal cost be given by $\overline{J}^{\pi} = \sum_{t=0}^T \overline{c}_t$, where $\overline{c}_t = c(\overline{\V{x}}_t,\overline{\V{u}}_t)$, for $t\leq T-1$, and $\overline{c}_T = c_T(\overline{\V{x}}_T,\overline{\V{u}}_T)$. Further, assume that the cost function is smooth enough that it permits the expansion $J^{\pi} = \overline{J} + \sum_t \M{C}_t \delta \V{x}_t + \sum_t \M{R}_t^c(\delta \V{x}_t)$ about the nominal trajectory, where $\M{C}_t$ denotes the linear term in the perturbation expansion and $\M{R}_t^c(\cdot)$ denote the second and higher order terms in the same. Finally, define the exactly linear perturbation system $\delta \V{x}_{t+1}^\ell = \M{A}_t \delta \V{x}_t^\ell + \M{B}_t\M{K}_t \delta \V{x}_t^\ell + \epsilon \M{B}_t \V{w}_t$. Further, let $\delta J_1^{\pi,\ell}$ denote the cost perturbation due to solely the linear system, i.e., $\delta J_1^{\pi,\ell} = \sum_t \M{C}_t \delta \V{x}_t^\ell$.  Then, the Decoupling result states the following \cite{D2C1.0}:
\begin{theorem}
The closed-loop cost function $J^{\pi}$can be expanded as $J^{\pi} = \overline{J}^{\pi} + \delta J_1^{\pi,\ell} + \delta J_2^{\pi}$. Furthermore, $\Exp{}{J^{\pi}} = \overline{J}^{\pi} + O(\epsilon^2)$, and $\Var[J^{\pi}] = \Var[\delta J_1^{\pi,\ell}] + O(\epsilon^4)$, where $\Var[\delta J_1^{\pi,\ell}]$ is $O(\epsilon^2)$.
\end{theorem}
Thus, the above result says the mean value of the cost is determined almost solely by the nominal control actions while the variance of the cost is almost solely determined by the linear closed-loop system. Thus, decoupling result says that the feedback law design can be decoupled into an open-loop and a closed-loop problem.\\
\textit{Open-Loop Problem:} This problem solves the deterministic/ nominal optimal control problem:
\begin{equation}
    J= \min_{\overline{\V{u}}_t} \sum_{t=0} ^{T-1} c(\overline{\V{x}}_t,\overline{\V{u}}_t) + c_T(\overline{\V{x}}_T),
\end{equation}
subject to the nominal dynamics: $\overline{\V{x}}_{t+1} = f(\overline{\V{x}}_t, \overline{\V{u}}_t)$. \\
\textit{Closed-Loop Problem:} One may try to optimize the variance of the linear closed-loop system
\begin{equation}
    \min_{\M{K}_t} \Var[\delta J_1^{\pi,\ell}]
\end{equation}
subject to the linear dynamics $\delta \V{x}_{t+1}^\ell = \M{A}_t \delta \V{x}_t^\ell + \M{B}_t \M{K}_t \delta \V{x}_t^\ell + \epsilon \M{B}_t \V{w}_t$.
However, the above problem does not have a standard solution but note that we are only interested in a good variance for the cost function and not the optimal one. Thus, this may be accomplished by a surrogate LQR problem that provides a good linear variance as follows.\\
\textit{Surrogate LQR Problem:} Here, we optimize the standard LQR cost:
\begin{equation}
    \delta J_{\textsc{lqr}} =\min_{\V{u}_t} \Exp{\V{w}_t}{\sum_{t=0}^{T-1} \delta {\tr{\V{x}}_t} \M{Q} \delta \V{x}_t + \delta \tr{\V{u}}_t\M{R}\delta \V{u}_t + \delta \tr{\V{x}}_T \M{Q}_f \delta \V{x}_T},
    \label{LQRcost}
\end{equation}
subject to the linear dynamics $\delta \V{x}_{t+1}^\ell = \M{A}_t \delta \V{x}_t^\ell + \M{B}_t \delta \V{u}_t + \epsilon \M{B}_t \V{w}_t$. In this paper, this decoupled design shall henceforth be called the trajectory-optimized LQR (T-LQR) design.  

\subsection{Multi-agent setting}
Now, we generalize the above result to a class of multi-agent problems. 
We consider a set of agents that are transition independent, i.e, their dynamics are independent of each other. For simplicity, we also assume that the agents have perfect state measurements. Let the system equations for the agents be given by:
$
\V{x}_{t+1}^j = f(\V{x}_t^j) + \M{B}^j_t(\V{u}_t^j + \epsilon \V{w}_t^j),
$
where $j = 1,2,\dots,M$ denotes the $j\Th$ agent. (We have assumed the control affine dynamics for simplicity). Further, let us assume that we are interested in the minimization of the joint cost of the agents given by $\mathcal{J} = \sum_{t=0}^{T-1} c(\M{X}_t,\M{U}_t) + \Phi(\M{X}_T)$, where $\M{X}_t = [\V{x}_t^1,\dots,\V{x}_t^M]$, and $\M{U}_t = [\V{u}_t^1,\dots, \V{u}_t^M]$  are the joint state and control action of the system. The objective of the multi-agent problem is minimize the expected value of the cost $\Exp{}{\mathcal{J}}$ over the joint feedback policy $\M{U}_t(\cdot)$. The decoupling result holds here too and thus the multi-agent planning problem can be separated into  an open and closed-loop problem. The open-loop problem consists of optimizing the joint nominal cost of the agents subject to the individual dynamics.\\
\textit{Multi-Agent Open-Loop Problem:}\\
\begin{align}\label{OL-MA}
\overline{\mathcal{J}} = \min_{\overline{\M{U}}_t} \sum_{t=0}^{T-1} c(\overline{\M{X}}_t,\overline{\M{U}}_t) + \Phi(\overline{\M{X}}_T), 
\end{align}
subject to the nominal agent dynamics
$
\overline{\V{x}}_{t+1}^j = f(\overline{\V{x}}_t^j) + \M{B}^j_t\overline{\V{u}}_t^j.
$
The closed-loop, in general, consists of optimizing the variance of the cost $\mathcal{J}$, given by $\Var[\delta \mathcal{J}^\ell_1]$, where $\delta \mathcal{J}_1^\ell = \sum_t \M{C}_t \delta \M{X}_t^l$ for suitably defined $\M{C}_t$, and $\delta \V{X}_t^\ell = [\delta \V{x}_t^1,\dots, \delta \V{x}_t^M]$, where the perturbations $\delta \V{x}_t^j$ of the $j\Th$ agent's state  is governed by the decoupled linear multi-agent system $\delta \V{x}_t^j = \M{A}_t\delta \V{x}_t^j + \M{B}_t^j \delta \V{u}^j_t + \epsilon \M{B}_t^j \V{w}_t^j.$ This design problem does not have a standard solution but recall that we are not really interested in obtaining the optimal closed-loop variance, but rather a good variance. Thus, we can instead solve a surrogate LQR problem given the cost function $\delta \mathcal{J}_{\textsc{mtlqr}} = \sum_{t=0}^{T-1} \sum_j \delta {\tr{\V{x}_t^j}} \M{Q}^j \delta \V{x}_t^j + \delta \tr{\V{u}_t^j}\M{R}\delta \V{u}_t^j + \sum_j\delta \tr{\V{x}_T^j} \M{Q}^j_f \delta \V{x}_T^j$. Since the cost function itself is decoupled, the surrogate LQR design degenerates into a decoupled LQR design for each agent.\\
\textit{Surrogate Decoupled LQR Problem:}
\begin{equation}
    \delta \mathcal{J}^j =\min_{\V{u}_t^j} \Exp{\V{w}_t^j}{\sum_{t=0}^{T-1} \delta {\tr{\V{x}_t^j}} \M{Q}^j \delta \V{x}_t^j + \delta \tr{\V{u}_t^j}\M{R}\delta \V{u}_t^j + \delta \tr{\V{x}_T^j} \M{Q}^j_f \delta \V{x}_T^j},
\end{equation}

subject to the linear decoupled agent dynamics $\delta \V{x}_t^j = \M{A}_t\delta \V{x}_t^j + \M{B}_t^j \delta \V{u}^j_t + \epsilon \M{B}_t^j \V{w}_t^j.$\\
\begin{remark}
Note that the above decoupled feedback design results in a spatial decoupling between the agents in the sense that, at least in the small noise regime, after their initial joint plan is made, the agents never need to communicate with each other in order to complete their missions.
\end{remark}

\subsection{Planning Complexity versus Uncertainty}
The decoupling principle outlined above shows that the complexity of planning can be drastically reduced while still retaining near optimal performance for sufficiently small noise (i.e., parameter  $\epsilon \ll 1$).  Nonetheless, the skeptical reader might argue that this result holds only for low values of $\epsilon$ and thus, its applicability for higher noise levels is suspect.  Still, because the result is second order, it hints that near optimality might be over a reasonably large $\epsilon$.  Naturally, the question is \textsl{`will it hold for medium to higher levels of noise?'}\\

\begin{figure}[t]
    \centering
    \begin{subfigure}[b]{0.225\textwidth}
        \centering
        \includegraphics[width=\textwidth]{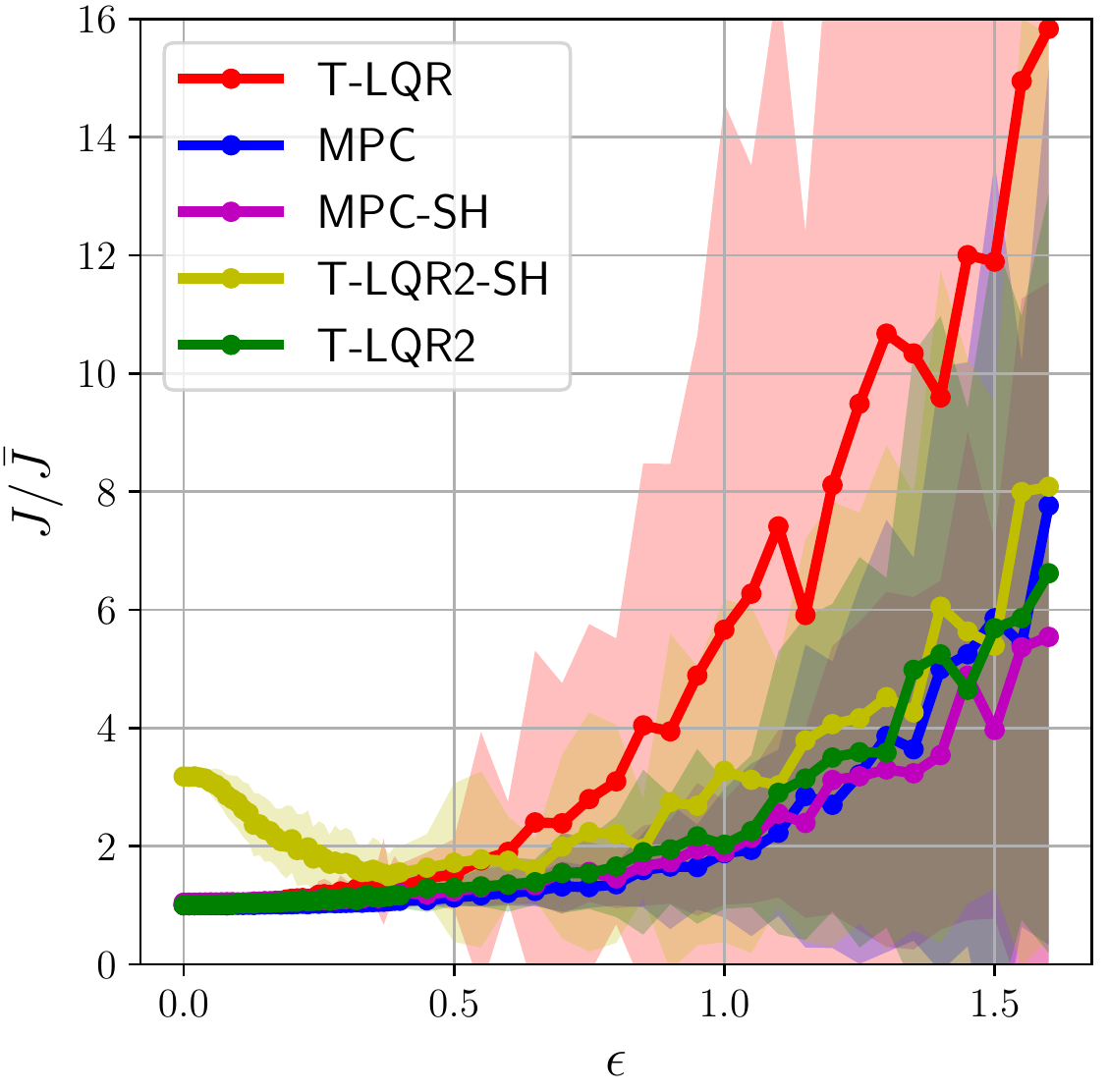}
        \caption{Full noise spectrum.}
        \label{1 agent cost full}
    \end{subfigure}%
    \begin{subfigure}[b]{0.225\textwidth}
        \centering
        \includegraphics[width=\textwidth]{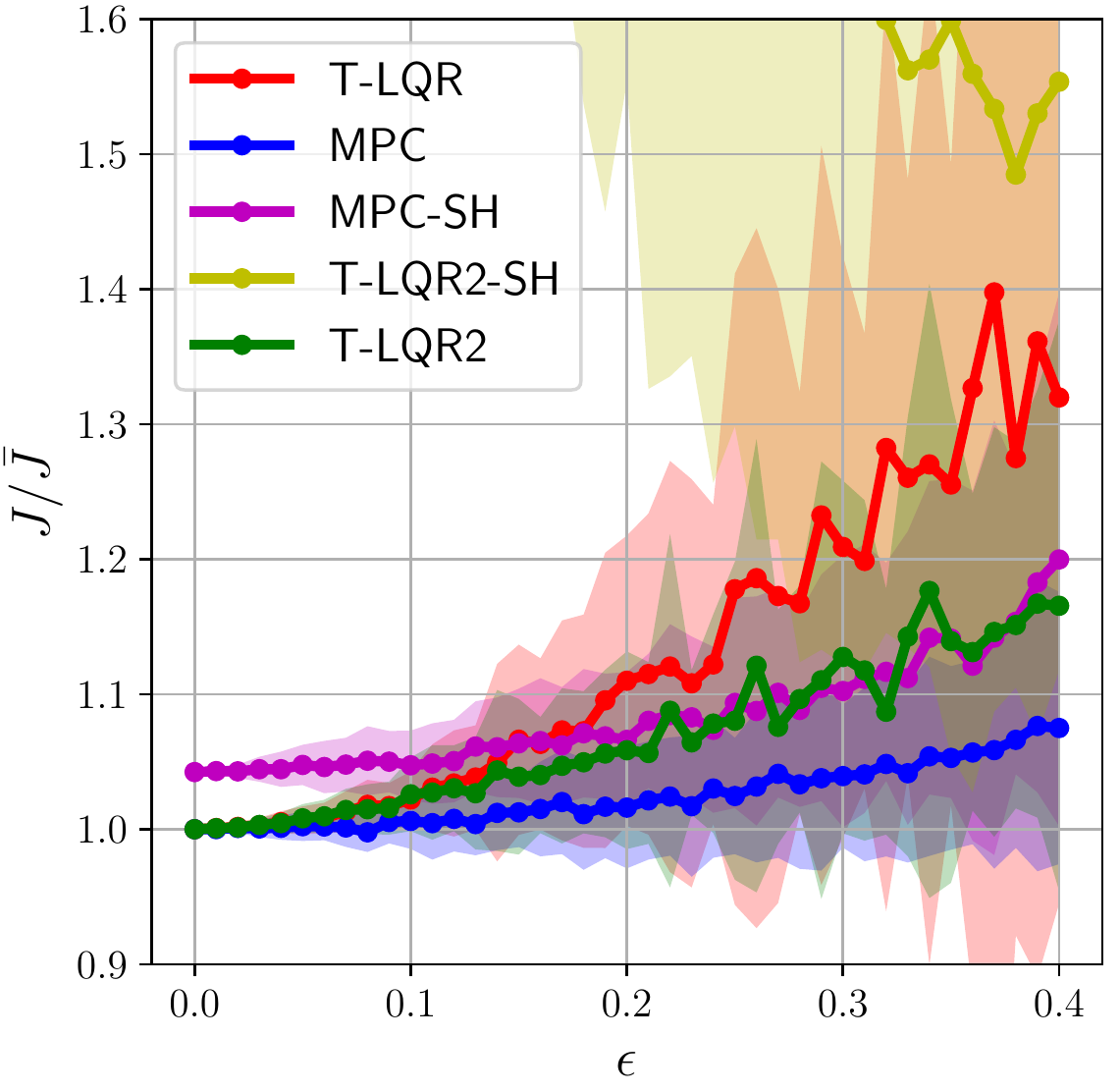}
        \caption{Enhanced detail: $0\leq\epsilon\leq0.4$.}
        \label{1 agent cost low}
    \end{subfigure}
    \caption{Cost evolution of the different algorithms for varying noise for a single agent. Control Horizon ($H_c$) used for MPC-SH and T-LQR2-SH was 7. $J_{\textrm{thresh}}$ = 2\% was the replanning threshold used. $J/\overline{J}$ is the ratio of the cost incurred during execution to the nominal cost and is used as the performance measure throughout the paper. The nominal cost $\overline{J}$ which is calculated by solving the deterministic OCP for the total time horizon, just acts as a normalizing factor here. (T-LQR2-SH is not shown in (b) since it skews the graph and is not important in low noise cases.)}
    \label{1_agent_cost}
\end{figure}%
\textit{Preview of our Results.} In this paper, we illustrate the
degree to which the above result still holds when we allow periodic replanning
of the nominal trajectory in T-LQR in an event triggered fashion, dubbed
T-LQR2.  Here, we shall use MPC as a ``gold standard'' for comparison since the
true stochastic control problem is intractable, and it was shown by Fleming in
a seminal paper~\cite{fleming1971stochastic} that, effectively speaking, the MPC policy is
$O(\epsilon^4)$ near-optimal compared to the true stochastic policy.  We show
that though the number of replanning operations in T-LQR2 increases the
planning burden over T-LQR, it is still much reduced when compared to MPC,
which replans continually.  The ability to trigger replanning means that T-LQR2
can always produce solutions with the same quality as MPC, albeit by demanding
the same computational cost as MPC in some instances. But for moderate levels
of noise, T-LQR2 can produce comparable quality output to MPC with substantial
computational savings.

In the high noise regime, replanning is more frequent but we
shall see that there is another consideration at play. Namely, that the
effective planning horizon decreases and there is no benefit in planning all
the way to the end rather than considering only a few steps ahead, and in fact, in some cases, it can harmful to consider the distant future. Noting that
as the planning horizon decreases, planning complexity decreases, this
helps recover tractability even in this regime.\\

Thus, while lower levels of noise render the planning problem tractable due to
the decoupling result, planning under even medium and higher levels of noise
can be practical because the planning horizon should shrink as uncertainty
increases.  When noise inundates the system, long-term predictions become so
uncertain that the best-laid plans will very likely run awry, then it would be wasteful to
invest significant time thinking very far ahead. 
To examine this widely-recognized truth more quantitatively,
the parameter $\epsilon$ will be a knob we adjust, exploring these
aspects in the subsequent analysis. 


%
\section{The Planning Algorithms}
The preliminaries and the algorithms are explained below:%
\subsection{Deterministic Optimal Control Problem:}
Given the initial state $\V{x}_0$ of the system, the solution to the deterministic OCP is given as:%
\begin{equation}
    J^{*}(\V{x}_0) = \min_{\V{u}_{0:T-1}} \left[\sum^{T-1}_{t=0} c_t(\V{x}_t, \V{u}_t) + c_T(\V{x}_T)\right], 
    \label{DOCP}
\end{equation}%
\begin{align*}
s.t. \ \V{x}_{t+1} = f(\V{x}_t) + \M{B}_t \V{u}_t,\\
   \V{u}_{\text{min}} \leq \V{u}_t \leq \V{u}_{\text{max}},\\
    | \V{u}_{t} - \V{u}_{t-1}| \leq \Delta \V{u}_{\text{max}}.
\end{align*}
The last two constraint model physical limits that impose upper bounds and lower bounds on control inputs and rate of change of control inputs. 
The solution to the above problem gives the open-loop control inputs $\overline{\V{u}}_{0:T-1}$ for the system.
For our problem, we take a quadratic cost function for state and control as 
$
    c_t(\V{x}_t,\V{u}_t) = \tr{\V{x}}_t  \M{W}^x\V{x}_t + \tr{\V{u}}_t\M{W}^u \V{u}_t,
$
$
    c_T(\V{x}_T) = \tr{\V{x}}_T\M{W}^x_f\V{x}_T,
$
where $\M{W}^x,\ \M{W}^x_f  \succeq \M{0}$ and $\M{W}^u \ \succ \ \M{0}$.\\

\subsection{Model Predictive Control (MPC):}
We employ the non-linear MPC algorithm due to the non-linearities associated
with the motion model. The MPC algorithm implemented here solves the
deterministic OCP~\eqref{DOCP} at every time step, applies the control inputs
computed for the first instant and uses the rest of the solution as an initial
guess for the subsequent computation. In the next step, the current state of
the system is measured and used as the initial state and the process is
repeated. 
\subsection{Short Horizon MPC (MPC-SH):}
We also implement a variant of MPC which is typically used in practical
applications where it solves the OCP only for a short horizon rather than the
entire horizon at every step. At the next step, a new optimization is solved
over the shifted horizon. This implementation gives a greedy solution but is
computationally easier to solve. It also has certain advantageous properties in
high noise cases which will be discussed in the results section. We denote the short planning horizon as $H_c$ also called as the control horizon, upto which the controls are computed. A generic algorithm for MPC is shown in
Algorithm~\ref{MPC_algo}. 
%
\begin{algorithm}[h]
\SetAlgoLined
    \KwIn{$\V{x}_0$ -- initial state, $\V{x}_g$ -- final state, $T$ -- time horizon, $H_c$ -- control horizon, $\Delta t$ -- time step, $\mathcal{P}$ -- system and environment parameters.}
    \For{$t \leftarrow 0$ \KwTo $T-1$}{
        $\V{u}_{t:t+H_c-1} \leftarrow$ \phantom{xxxxxxx}OCP($\V{x}_{t},\V{x}_g, \min(H_c, T\!-\!t),\V{u}_{t-1}, \V{u}_{\textrm{guess}},\mathcal{P}$)
        $\V{x}_{t+1} \leftarrow$ $f(\V{x}_{t}) + \M{B}_t(\V{u}_t + \epsilon\V{w}_t)$ 
    }
    \caption{MPC algorithm\label{MPC_algo}.}
\end{algorithm}

\subsection{Trajectory Optimised Linear Quadratic Regulator~\mbox{(T-LQR)}:}
\label{sec:lqr_gains}

As discussed in Section~\ref{section:decoupling}, stochastic optimal control problem can be decoupled and solved by designing an optimal open-loop (nominal) trajectory and a decentralized LQR policy to track the nominal. 
\\
\textit{Design of nominal trajectory}: The nominal trajectory is generated by first finding the optimal open-loop control sequence by solving the deterministic OCP~\eqref{DOCP} for the system. Then, using the computed control inputs and the noise-free dynamics, the sequence of states traversed $\overline{\V{x}}_{0:T}$ can be calculated.\\
\textit{Design of feedback policy:} In order to design the LQR controller, the system is first linearised about the nominal trajectory ($\overline{\V{x}}_{0:T}$, $\overline{\V{u}}_{0:T-1}$). Using the linear time-varying system, the feedback policy is determined by minimizing a quadratic cost as shown in~\eqref{LQRcost}. %
The linear quadratic stochastic control problem~\eqref{LQRcost} can be easily solved using the algebraic Riccati equation and the resulting policy is $\delta{\V{u}}_{t} = -\M{L}_t\delta{\V{x}}^\ell_t$. The feedback gain and the Riccati equations are given by

\begin{equation}
    \M{L}_t = \inv{(\M{R} + \M{B}^T_t\M{P}_{t+1} \M{B}_t)} \tr{\M{B}}_t \M{P}_{t+1}  \M{A}_t,
    \label{LQR_gain}
\end{equation}

\begin{equation}
    \M{P}_{t} = \tr{\M{A}}_t \M{P}_{t+1} \M{A}_t - \tr{\M{A}}_t \M{P}_{t+1} \M{B}_t  \M{L}_t +  \M{Q}, 
    \label{Riccati}
\end{equation}
respectively where $\M{Q}_f, \M{Q} \succeq \V{0}, \M{R} \ \succ \V{0}$ are the weight matrices for states and control. Here~\eqref{Riccati} is the discrete-time dynamic Riccati
equation which can be solved by backward iteration using the terminal condition
$\M{P}_{T} = \M{Q}_f $.

\subsection{T-LQR with Replanning (\mbox{T-LQR2}):}
T-LQR performs well at low noise levels, but at medium and high noise levels the system tends to deviate from the nominal. So, at any point during the execution if the deviation is beyond a threshold $J_{\textrm{thresh}}= \frac{J_t - \overline{J}_t}{\overline{J}_t}$, where $J_t$ denotes the actual cost during execution till time $t$ while $\overline{J}_t$ denotes the nominal cost. The factor $J_{\textrm{thresh}}$ measures the percentage deviation of the online trajectory from the nominal, and replanning is triggered for the system from the current state for the remainder of the horizon. Note that if we set $J_{\textrm{thresh}}=0$, T-LQR2 reduces to MPC. The calculation of the new nominal trajectory and LQR gains are carried out similarly to the explaination in Section~\ref{sec:lqr_gains}. A generic algorithm for T-LQR and T-LQR2 is shown in Algorithm~\ref{TLQR_algo}.
\subsection{Short Horizon T-LQR with Replanning (T-LQR2-SH):}
A T-LQR equivalent of MPC-SH is also implemented where the nominal is planned only for a short horizon and it is tracked with a feedback policy as described in T-LQR. It also inherits the replanning property of T-LQR2.\\
The implementations of all the algorithms are available at \url{https://github.com/MohamedNaveed/Stochastic_Optimal_Control_algos/}.

\begin{algorithm}[h]
\SetAlgoLined
\SetKwProg{Fn}{Function}{ is}{end}
\SetKwComment{Comment}{/*}{*/}   
    \KwIn{$\V{x}_0$ -- initial state, $\V{x}_g$ -- final state, $T$ -- time horizon, $J_{\textrm{thresh}}$ -- replan threshold, $\Delta t$ -- time step, $\mathcal{P}$ -- system and environment parameters.}
    
    \Fn{Plan($\V{x}_{0},\V{x}_g, T, \V{{u}}_{\textrm{init}}$, $\V{u}_{\textrm{guess}}$,$\mathcal{P}$)}{
    $\overline{\V{{u}}}_{0:T-1}$ $\leftarrow$ OCP($\V{x}_{0},\V{x}_g, T,\V{{u}}_{\textrm{init}}$, $\V{u}_{\textrm{guess}}$,$\mathcal{P}$)
    
    \For{$t \leftarrow 0$ \KwTo $T-1$}{
        $\overline{\V{x}}_{t+1} \leftarrow f(\overline{\V{x}}_{t}) + \M{B}_t\overline{\V{u}}_t$
    }
    $\M{L}_{0:T-1} \leftarrow $ $Compute\_LQR\_Gain(\overline{\V{x}}_{0:T-1},\overline{\V{u}}_{0:T-1}$)
    
    return $\overline{\V{x}}_{0:T},\overline{\V{u}}_{0:T-1},\M{L}_{0:T-1}$ 
    }
    \Fn{Main()}{
    $\overline{\V{x}}_{0:T}$,$\overline{\V{u}}_{0:T-1}$,$\M{L}_{0:T-1}$ $\leftarrow$ $\textrm{Plan}(\V{x}_{0},\V{x}_g, T,\mathbf{0}, \V{u}_{\textrm{guess}},\mathcal{P})$
    
    \For{$t \leftarrow 0$ \KwTo $T-1$}{
        
        $\V{u}_{t} \leftarrow \overline{\V{u}}_{t} - \M{L}_{t}(\V{x}_{t} - \overline{\V{x}}_{t})$
        
        $\V{u}_{t} \leftarrow \textrm{Constrain}(\V{u}_{t})$ \tcp*[f]{ Enforce limits}
        
        $\V{x}_{t+1} \leftarrow f(\V{x}_{t}) + \M{B}_t(\V{u}_t + \epsilon\V{w}_t)$ 
        
        \If(\tcp*[f]{Replan?}){$ (J_t - \overline{J}_t)/\overline{J}_t  > J_{\textrm{thresh}}$}{
            $\overline{\V{x}}_{t+1:T}, \overline{\V{u}}_{t+1:T-1}, \M{L}_{t+1:T-1} \leftarrow \textrm{\phantom{xxxxxx}Plan}(\V{x}_{t+1},\V{x}_g, T\!-\!t\!-\!1,\V{u}_t, \V{u}_{\textrm{guess}},\mathcal{P})$
            
        }
    }
    }
    \caption{T-LQR algorithm with replanning}\label{TLQR_algo}
\end{algorithm}

\subsection{Multi-Agent versions}
\begin{figure}[t]
    \centering
    \begin{subfigure}[b]{0.225\textwidth}
        \centering
        \includegraphics[width=\textwidth]{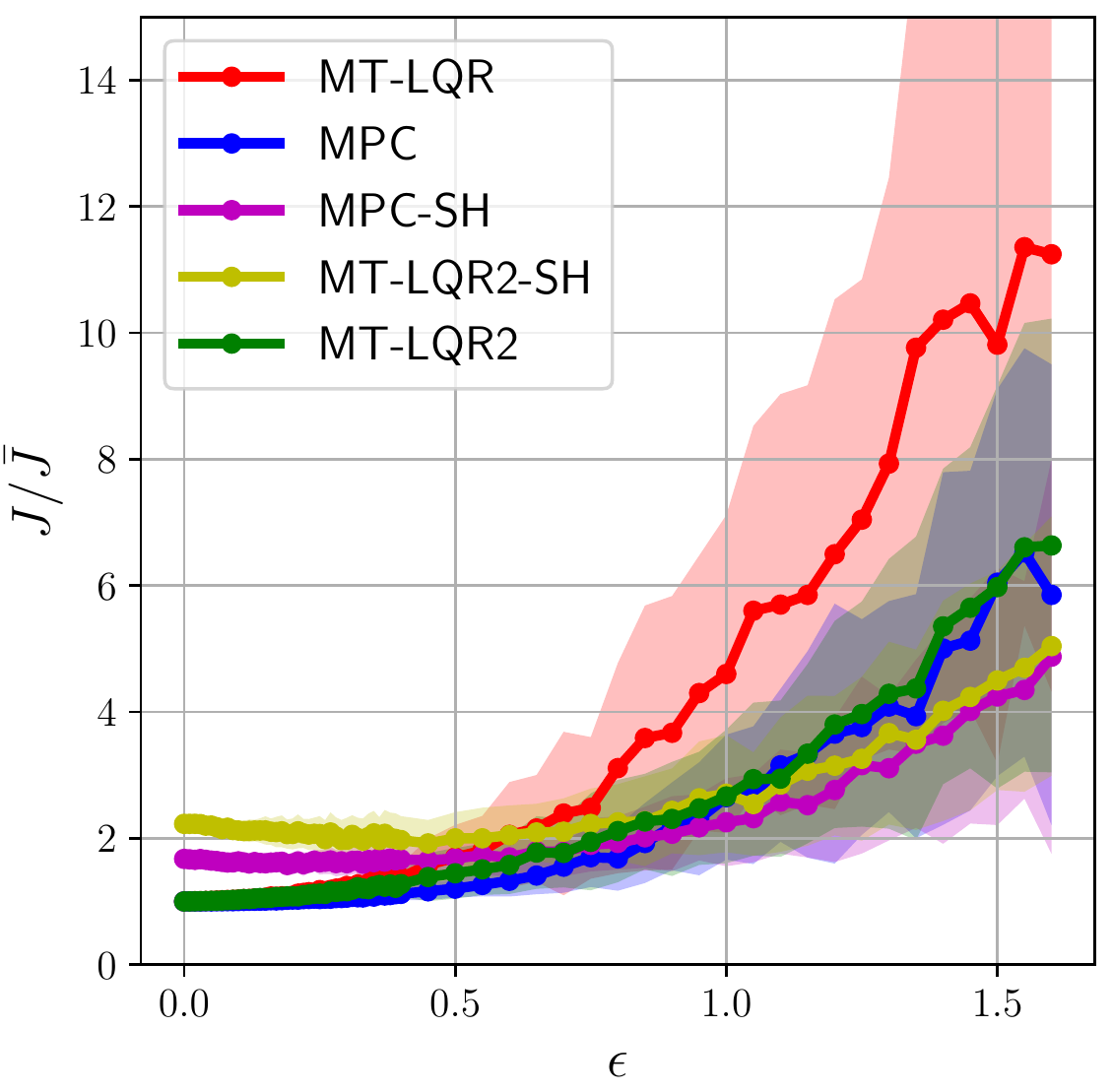}
        \caption{Full noise spectrum.}
        \label{3 agent cost full}
    \end{subfigure}
    \begin{subfigure}[b]{0.225\textwidth}
        \centering
        \includegraphics[width=\textwidth]{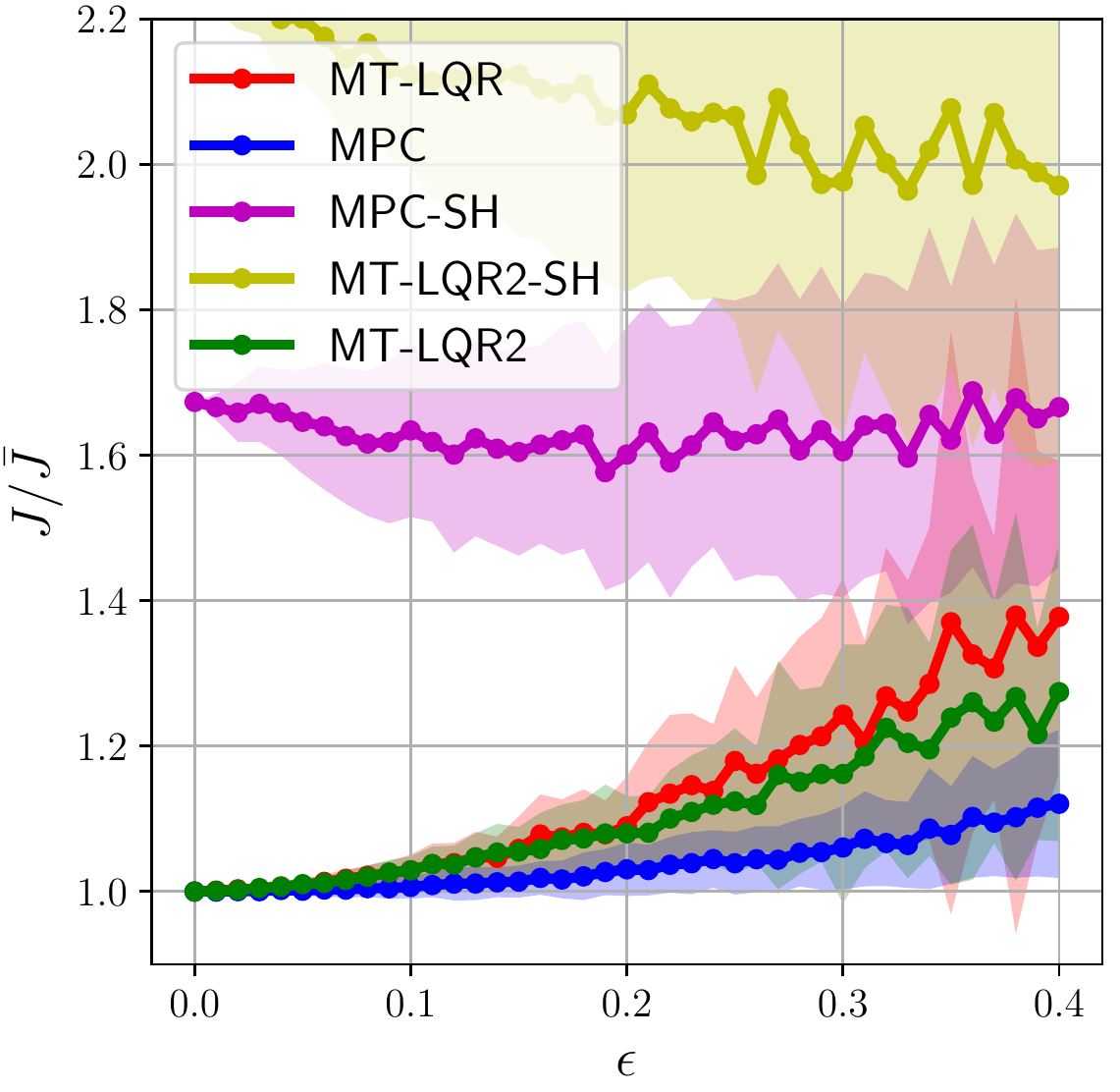}
        \caption{Enhanced detail: $0\leq\epsilon\leq0.4$.}
        \label{3 agent cost low}
    \end{subfigure}
     \caption{Cost evolution of the different algorithms for varying noise for 3 agents. Control Horizon ($H_c$) used for MPC-SH and MT-LQR2-SH was 7. $J_{\textrm{thresh}}$ = 2\% was the replanning threshold used.}
    \label{3_agent_cost}
\end{figure}%
The MPC version of the multi-agent planning problem is reasonably straightforward except that the complexity of the planning increased (exponentially) in the number of agents. Also, we note that the agents have to always communicate with each other in order to do the planning.\\
The Multi-agent Trajectory-optimised LQR (MT-LQR) version is also relatively straightforward in that the agents plan the nominal path jointly once, and then the agents each track their individual paths using their decoupled feedback controllers. There is no communication whatsoever between the agents during this operation.\\

The MT-LQR2 version is a little more subtle. The agents have to periodically
replan when the total cost deviates more than $J_{\textrm{thresh}}$ away from the
nominal, i.e., the agents do not communicate until the need to replan arises.
In general, the system would need to detect this in a distributed fashion, and
trigger replanning. We postpone consideration of this aspect of the problem to a
subsequent paper more directly focused on networking considerations. We will
assume that there exists a (yet to be determined) distributed strategy that
would perform the detection and replanning.

\subsection{Analysis of the High Noise Regime}
In this section, we perform a rudimentary analysis of the high noise regime. The medium noise case is more difficult to analyze and is left for future work, along with a more sophisticated treatment of the high noise regime.\\
First, recall the Dynamic Programming (DP) equation for the backward pass to determine the optimal time varying feedback policy:
$
    J_t(\V{x}_t) = \min_{\V{u}_t}\left\{c(\V{x}_t,\V{u}_t) + \Exp{}{J_{t+1}(\V{x}_{t+1})}\right\},
$
where $J_t(\V{x}_t)$ denotes the cost-to-go at time $t$ given the state is $\V{x}_t$, with the terminal condition $J_T(\cdot) = c_T(\cdot)$ where $c_T$ is the terminal cost function, and the next state $\V{x}_{t+1} = f(\V{x}_t) + \M{B}_t(\V{u}_t + \epsilon \V{w}_t)$. Suppose now that the noise is so high that $\V{x}_{t+1} \approx \M{B}_t \epsilon \V{w}_t$, i.e., the dynamics are completely swamped by the noise.\\
Consider now the expectation $\Exp{}{c_T(\V{x}_{t+1})}$ given some control $\V{u}_t$ was taken at state $\V{x}_t$. Since $\V{x}_{t+1}$ is determined entirely by the noise, $\Exp{}{c_T(\V{x}_{t+1})} = \int c_T(\M{B}_t\epsilon \V{w}_t)\mathbf{p}(\V{w}_t) d\V{w}_t = \overline{c_T}$, where $\overline{c_T}$ is a constant regardless of the previous state and control pair $\V{x}_t, \V{u}_t$. This observation holds regardless of the function $c_T(\cdot)$ and the time $t$.\\
Next, consider the DP iteration at time $T-1$. Via the argument above, it follows that $\Exp{}{J_T(\V{x}_T)}=\Exp{}{c_T(\V{x}_T)} = \overline{c_T}$, regardless of the state control pair $\V{x}_{T-1},\V{u}_{T-1}$ at the $(T-1)^{th}$ step, 
and thus, the minimization reduces to 
$J_{T-1}(\V{x}_{T-1}) = \min_{\V{u}} \left\{c(\V{x}_{T-1},\V{u})  + \overline{c_T}\right\}$, 
and thus, the minimizer is just the greedy action $\V{u}^*_{T-1} = \argmin_{\V{u}} c(\V{x}_{T-1},\V{u})$ due to the constant bias $\overline{c_T}$. 
The same argument holds for any $t$ since, although there might be a different $J_{t}(\cdot)$ at every time $t$, the minimizer is still the greedy action that minimizes $c(\V{x}_t,\V{u})$ as the cost-to-go from the next state is averaged out to simply some $\bar{J}_{t+1}$.\\


\begin{figure}
    \begin{subfigure}[b]{0.25\textwidth}
        \centering
        \includegraphics[width=\textwidth]{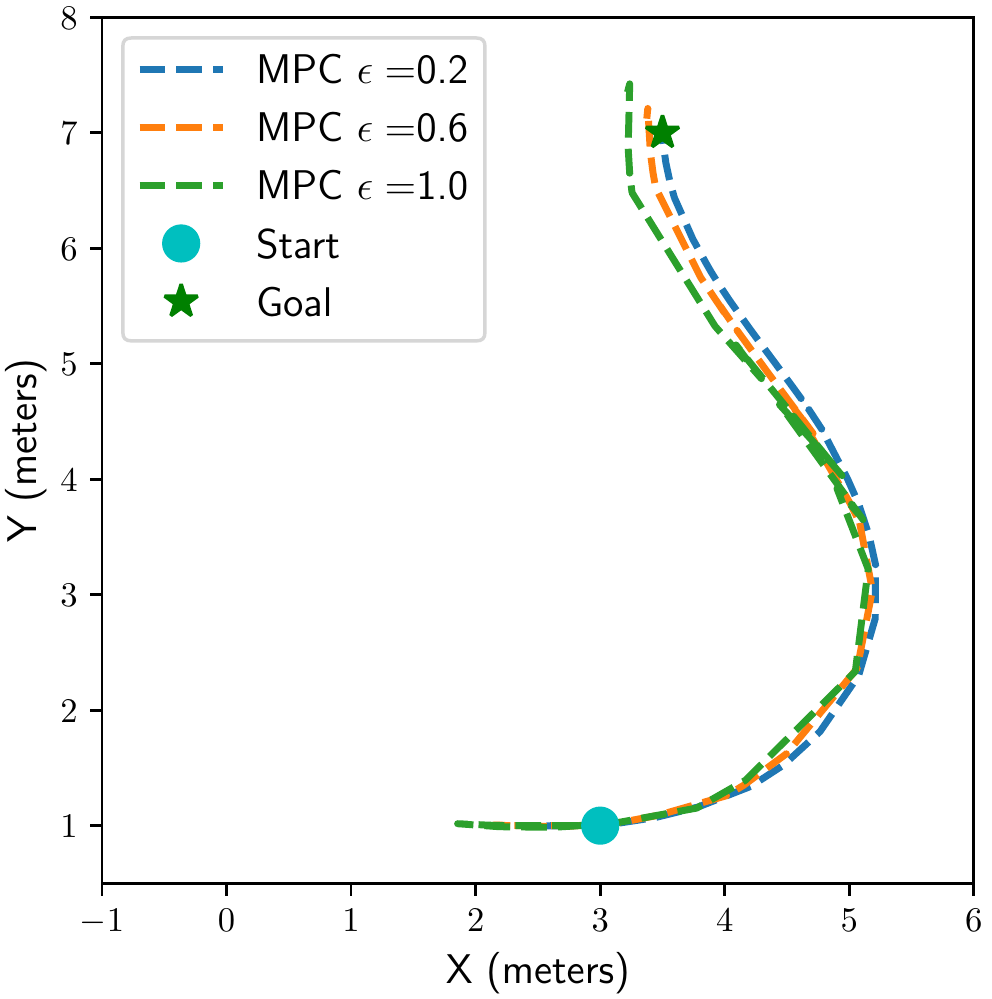}
        \label{test_cases_high_mpc}
        \caption{MPC}
    \end{subfigure}%
    \begin{subfigure}[b]{0.25\textwidth}
        \centering
        \includegraphics[width=\textwidth]{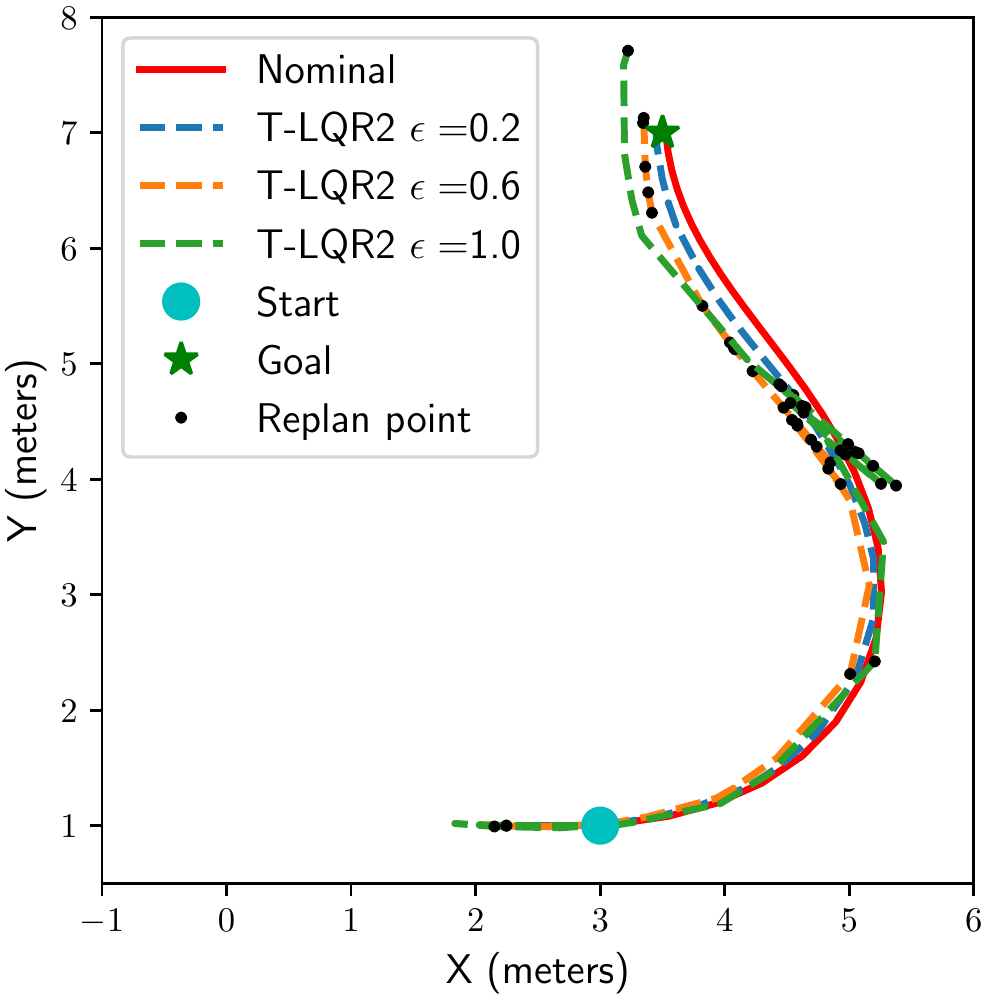}
        \label{test_cases_high_tlqr_replan}
        \caption{T-LQR2}
    \end{subfigure}
    \newline
    \begin{subfigure}[b]{0.25\textwidth}
        \centering
        \includegraphics[width=\textwidth]{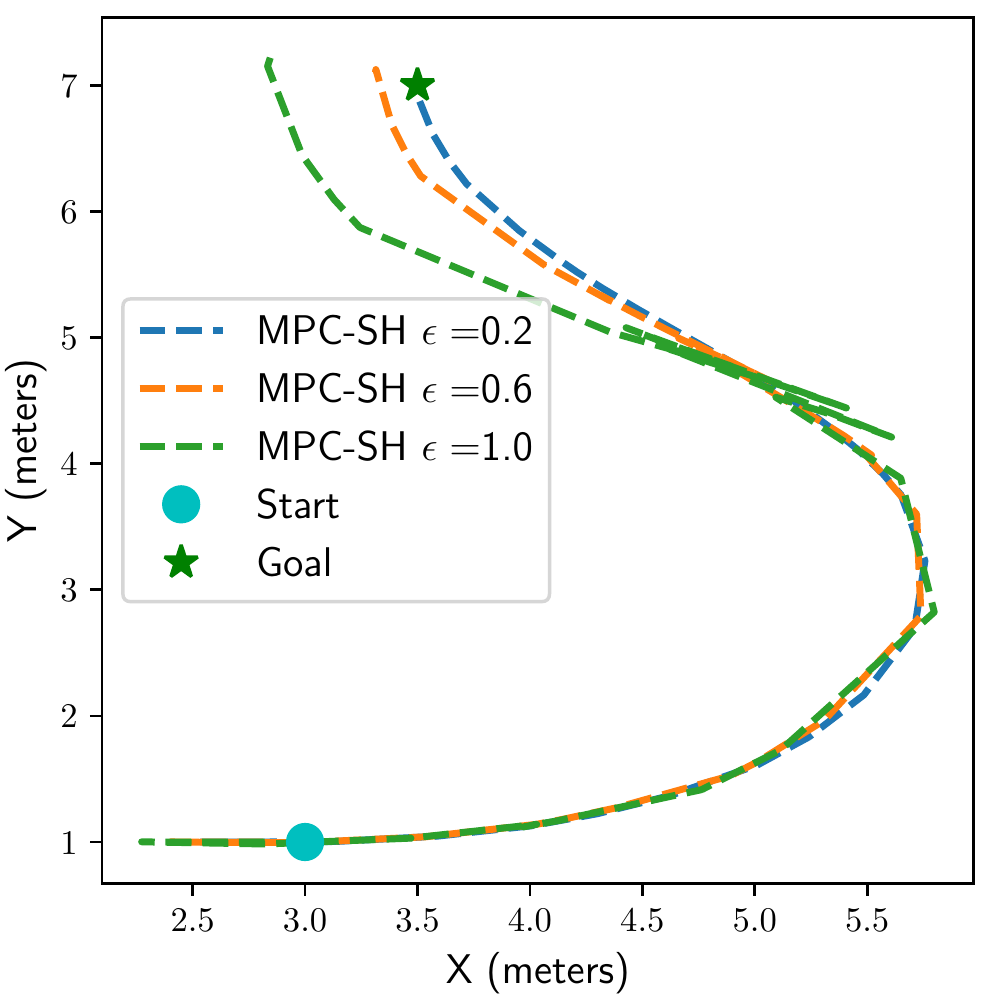}
        \label{test_cases_high_shmpc}
        \caption{MPC-SH}
    \end{subfigure}%
    \begin{subfigure}[b]{0.25\textwidth}
        \centering
        \includegraphics[width=\textwidth]{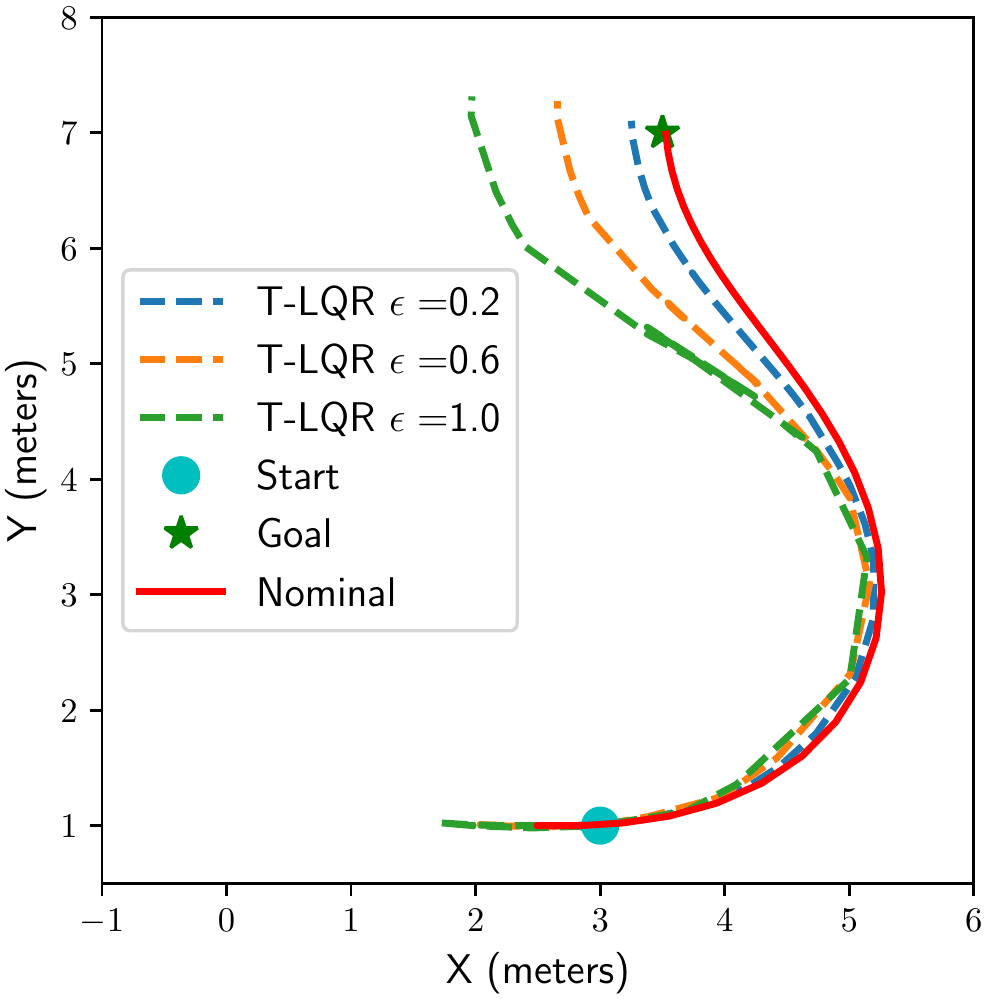}
        \label{test_cases_high_tlqr}
        \caption{T-LQR}
    \end{subfigure}
    \caption{Performance of the algorithms for varying levels of noise.}
    \label{fig:test_cases_high}
\end{figure} 

\section{Simulation Results:}
\begin{figure}[h]
    \begin{subfigure}[b]{0.25\textwidth}
        \centering
        \includegraphics[width=\textwidth]{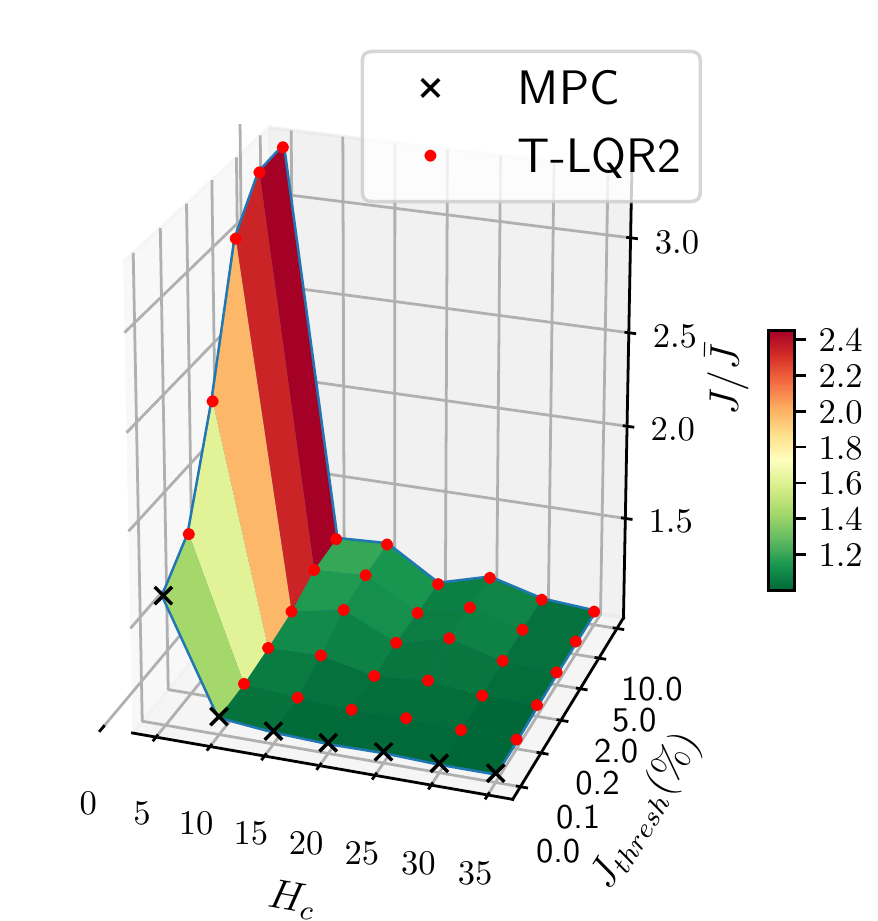}
        \caption{}
        \label{costvsHc_1_1}
    \end{subfigure}%
    \begin{subfigure}[b]{0.25\textwidth}
        \centering
        \includegraphics[width=\textwidth]{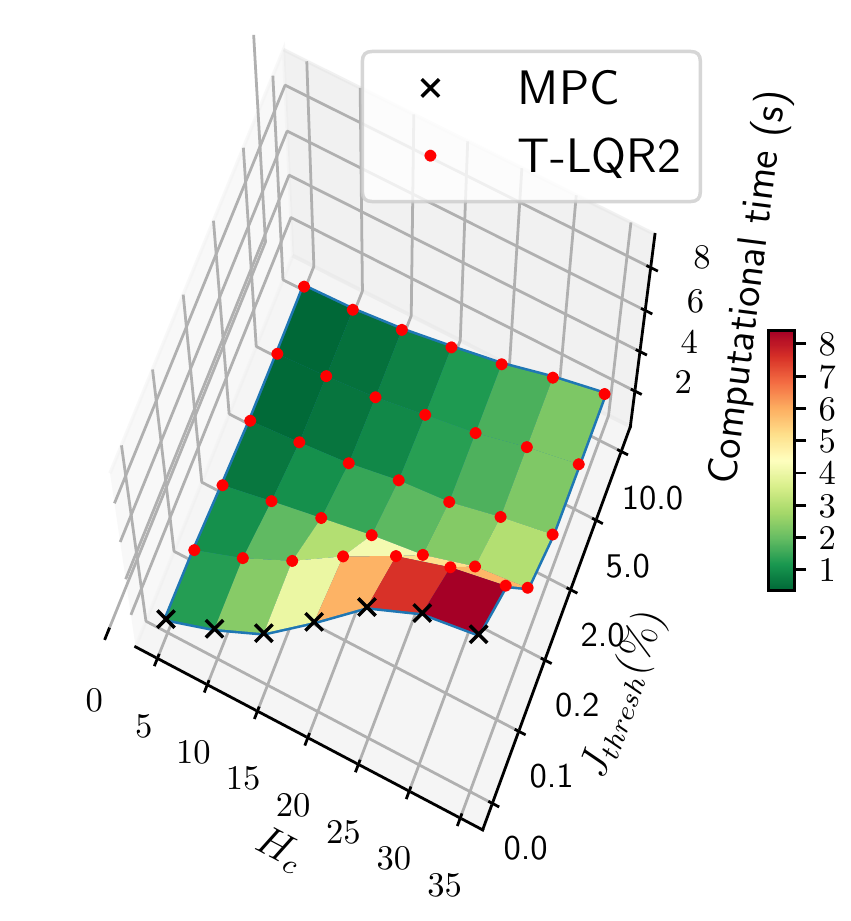}
        \caption{}
        \label{timevsHc_1_1}
    \end{subfigure}%
    \newline
    \begin{subfigure}[b]{0.25\textwidth}
        \centering
        \includegraphics[width=\textwidth]{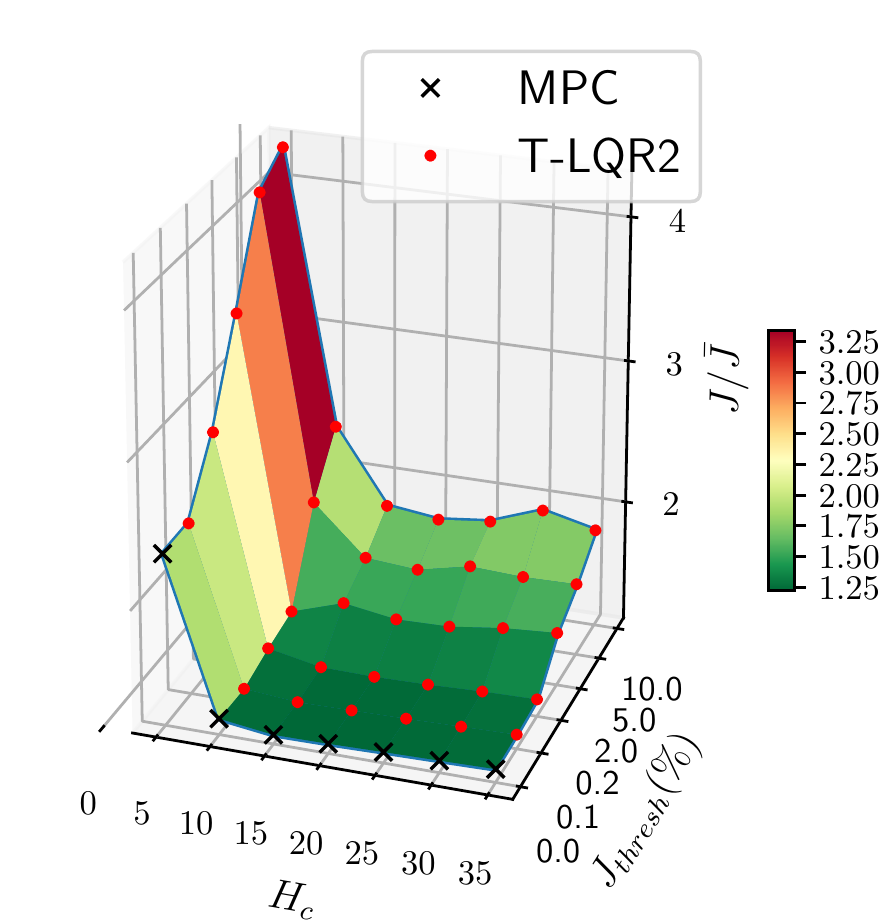}
        \caption{}
        \label{costvsHc_1_7}
    \end{subfigure}%
    \begin{subfigure}[b]{0.25\textwidth}
        \centering
        \includegraphics[width=\textwidth]{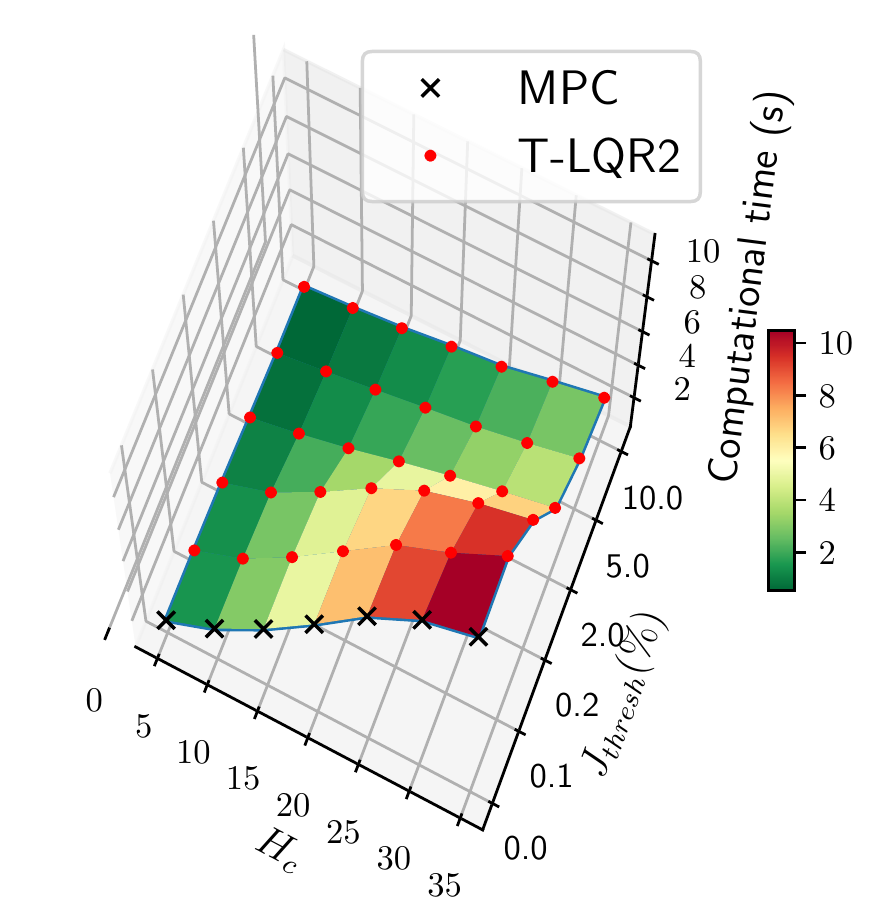}
        \caption{}
        \label{timevsHc_1_7}
    \end{subfigure}%
    \caption{Variation seen in cost incurred and computation time by changing the $J_{\textrm{thresh}}$ and control horizon ($H_c$) in T-LQR2 and MPC for a single agent case. (a) and (b) show the performance in terms of cost and computation time respectively for the same experiment at $\epsilon = 0.1$. Similarly, (c) and (d) show for $\epsilon=0.7$. Though MPC doesn't have a threshold for replanning, it is plotted at $J_{\textrm{thresh}} = 0\%$ since it replans at every time step.} 
    \label{fig:1_agent_3d}
\end{figure}

\begin{figure}[h]
    \begin{subfigure}[b]{0.25\textwidth}
        \centering
        \includegraphics[width=\textwidth]{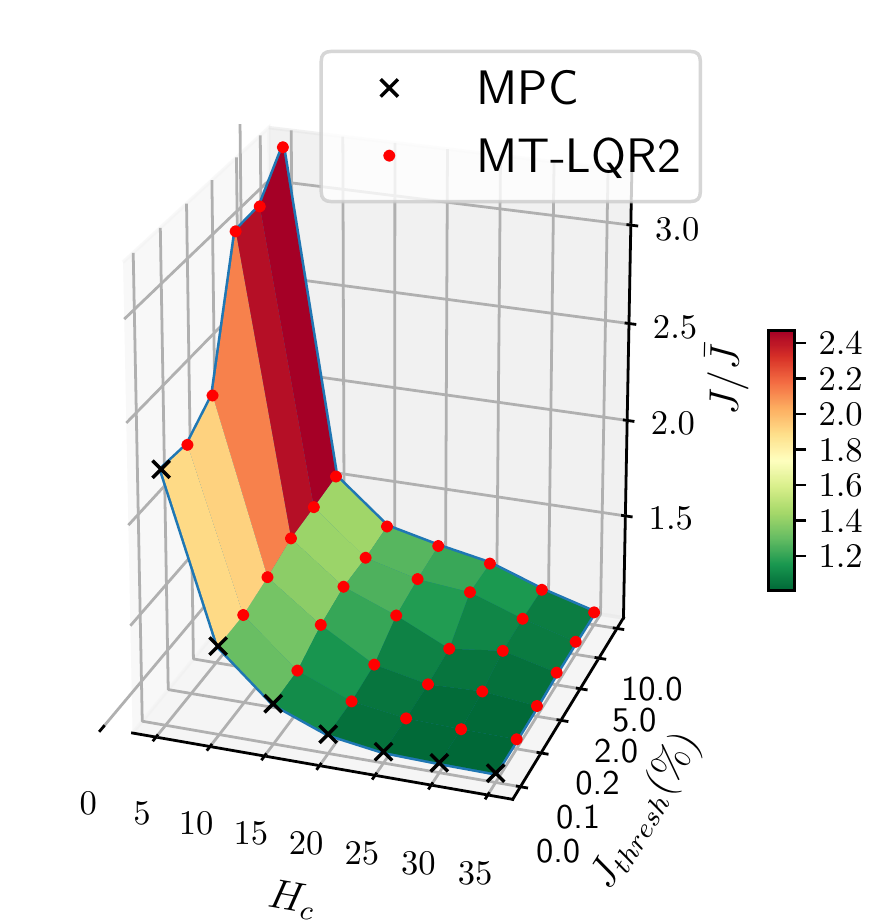}
        \caption{}
        \label{fig:costvsHc_3_1}
    \end{subfigure}%
    \begin{subfigure}[b]{0.25\textwidth}
        \centering
        \includegraphics[width=\textwidth]{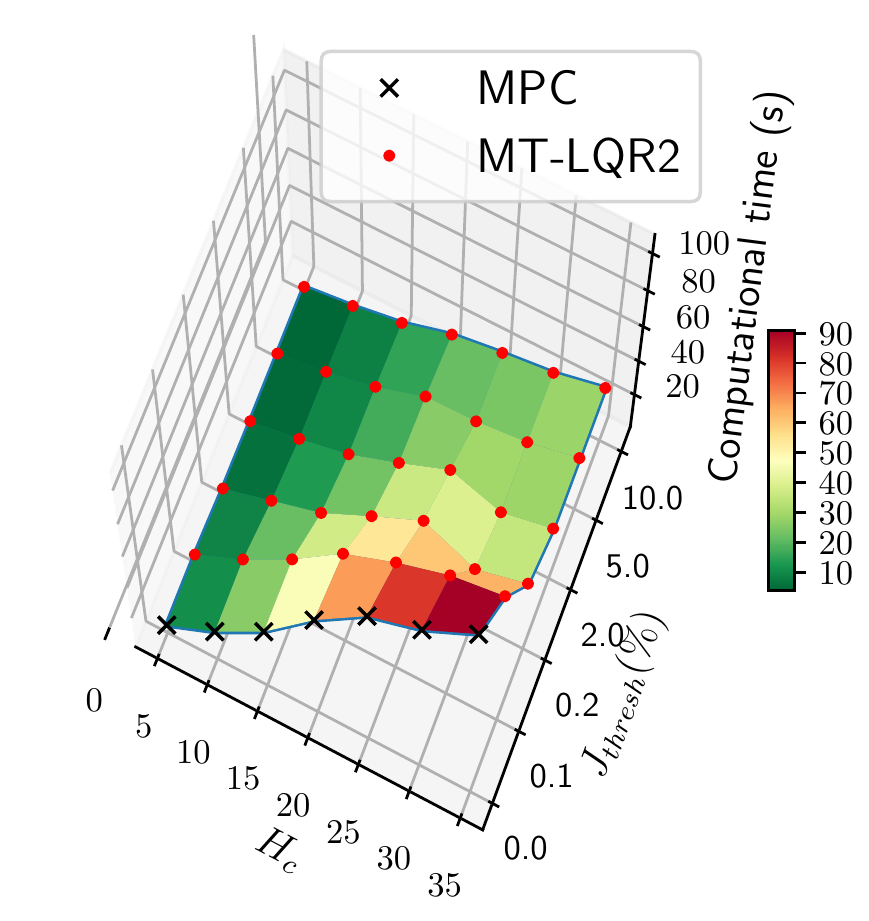}
        \caption{}
        \label{fig:timevsHc_3_1}
    \end{subfigure}%
    \newline
    \begin{subfigure}[b]{0.25\textwidth}
        \centering
        \includegraphics[width=\textwidth]{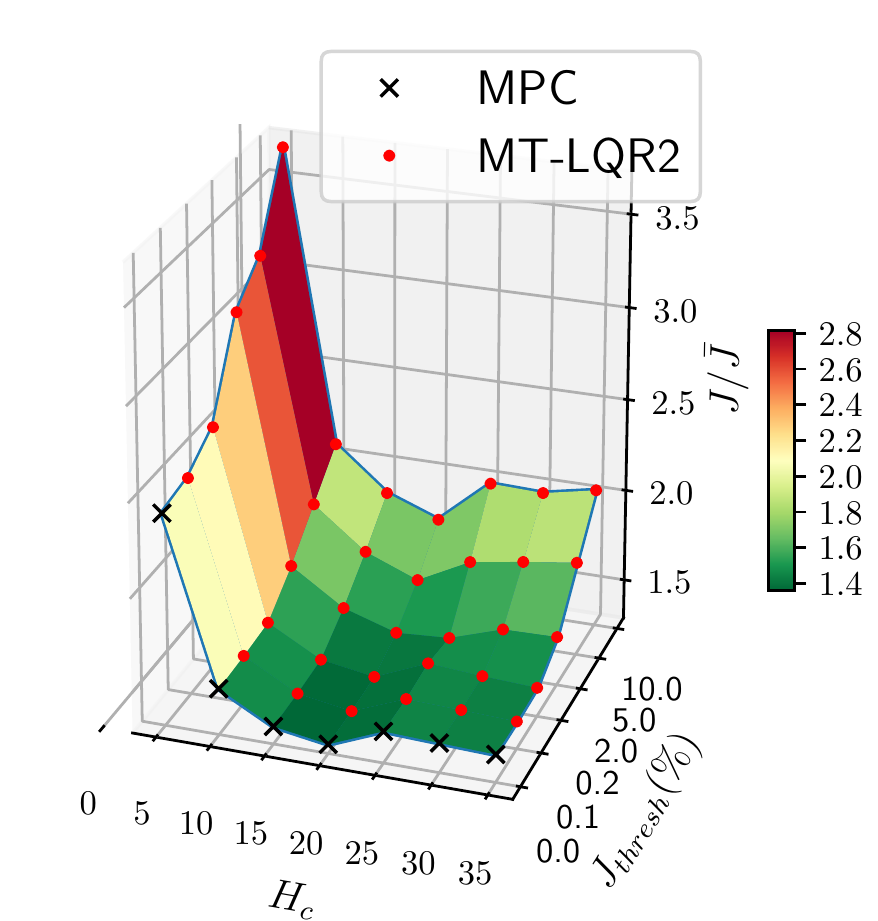}
        \caption{}
        \label{fig:costvsHc_3_7}
    \end{subfigure}%
    \begin{subfigure}[b]{0.25\textwidth}
        \centering
        \includegraphics[width=\textwidth]{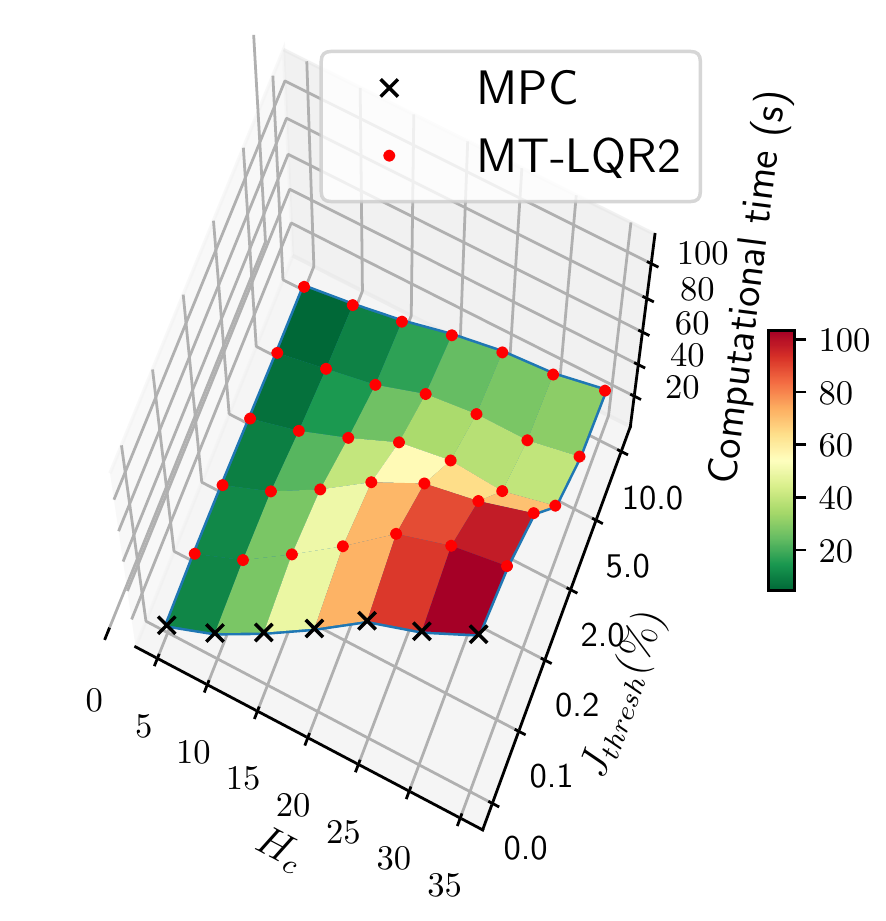}
        \caption{}
        \label{fig:timevsHc_3_7}
    \end{subfigure}%
    \caption{Variation seen in cost incurred and computational time by changing the $J_{\textrm{thresh}}$ and control horizon ($H_c$) in MT-LQR2 and MPC for 3 agents. (a) and (b) show for $\epsilon = 0.1$, (c) and (d) show for $\epsilon=0.7$.}
    \label{fig:3_agent_3d}
\end{figure}

%
%
    
%

%
We test the performance of the algorithms in a car-like robot model. Numerical optimization is carried out using \texttt{CasADi} framework \cite{Andersson2018} with \texttt{Ipopt} \cite{Ipopt} NLP solver in \texttt{Python}. To provide a good estimate of the performance the results presented were averaged from 100 simulations for every value of noise considered. Simulations were carried out in parallel across 100 cores in a cluster equipped with Intel Xeon 2.5GHz E5-2670 v2 10-core processors. The experiments chosen were done with a time horizon T = 35.

\subsection*{Car-like robot model:}
The car-like robot considered in our work has the following motion model:
\begin{align*}
    x_{t+1} &= x_t + v_{t}\cos(\theta_t)\Delta t,  &  \theta_{t+1} &= \theta_{t} + \frac{v_t}{L}\tan(\phi_t)\Delta t, \\
    y_{t+1} &= y_t + v_{t}\sin(\theta_t)\Delta t,  &  \phi_{t+1} &= \phi_{t} + \omega_t \Delta t,
\end{align*}

where $\tr{(x_t, y_t, \theta_t, \phi_t)}$ denote the robot's state vector
namely, robot's $x$ and $y$ position, orientation and steering angle at time $t$.
Also, $\tr{(v_t, \omega_t)}$ is the control vector and denotes the robot's linear
velocity and angular velocity (i.e., steering). Here $\Delta t$ is the
discretization of the time step. The values of the parameters used in the
simulation were $L = \SI{0.5}{\meter}$ and $\Delta t = \SI{0.1}{\second}$.

\subsection*{Noise characterization:}
We add zero mean independent identically distributed (i.i.d), random sequences ($\V{w}_t$) as actuator noise to test the performance of the control scheme. The standard deviation of the noise is $\epsilon$ times the maximum value of the corresponding control input, where $\epsilon$ is a scaling factor which is varied during testing, that is:
$
\V{w}_t = \V{u}_{\textrm{max}} \bm{\nu};  \quad \bm{\nu} \sim \mathcal{N}(\V{0}, \M{I}) 
$
and the noise is added as $\epsilon \V{w}_t$. Note that, we enforce the constraints in the control inputs before the addition of noise, so the controls can even take a value higher after noise is added.

\subsection{Single agent setting:}
A car-like robot is considered and is tasked to move from a given initial pose
to a goal pose. The environment of the robot is shown in
Figure~\ref{fig:test_cases_high}. The experiment is done for all the control
schemes discussed and their performance for different levels of noise are shown
in Figure~\ref{1_agent_cost}.

\subsection{Multi-agent setting:}
A labelled point-to-point transition problem with 3 car-like robots is considered where each agent is assigned a fixed destination which cannot be exchanged with another agent. The performance of the algorithms is shown in Figure~\ref{3_agent_cost}. The cost function involves the state and control costs for the entire system similar to the single agent case. One major addition to the cost function is the penalty function to avoid inter-agent collisions which is given by 
$
    \Psi^{(i,j)} = \textrm{M}\exp\left(-(\Vert \V{p}_t^i - \V{p}_t^j\Vert_2^2 - r_{\textrm{thresh}}^2)\right) 
$
where $\textrm{M} > 0$ is a scaling factor, $\V{p}^i_t = (x^i_t, y^j_t)$ and $r_{\textrm{thresh}}$ is the desired minimum distance the agents should keep between themselves.    

\begin{figure}[h]
    \centering
    \begin{subfigure}[b]{0.225\textwidth}
        \centering
        \includegraphics[width=\textwidth]{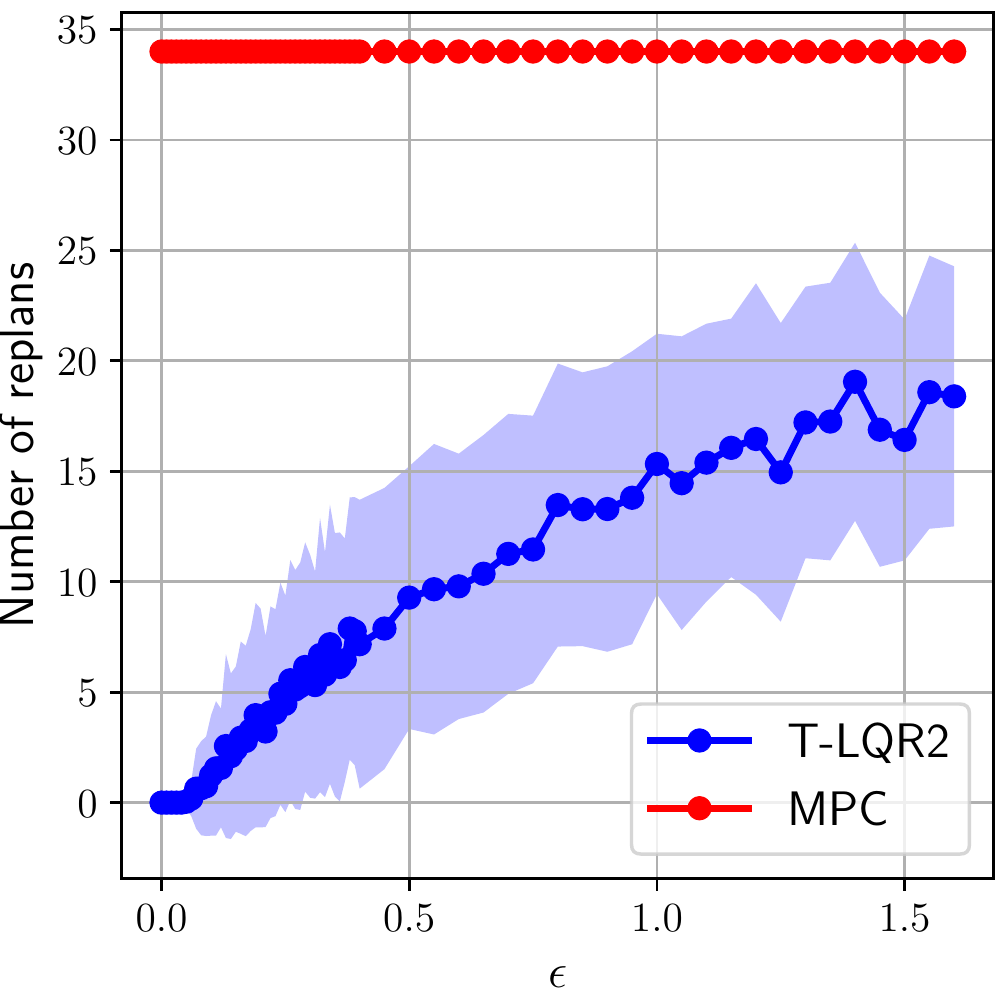}
        \caption{A single agent.}
        \label{1_agent_replan}
    \end{subfigure}
    \begin{subfigure}[b]{0.225\textwidth}
        \centering
        \includegraphics[width=\textwidth]{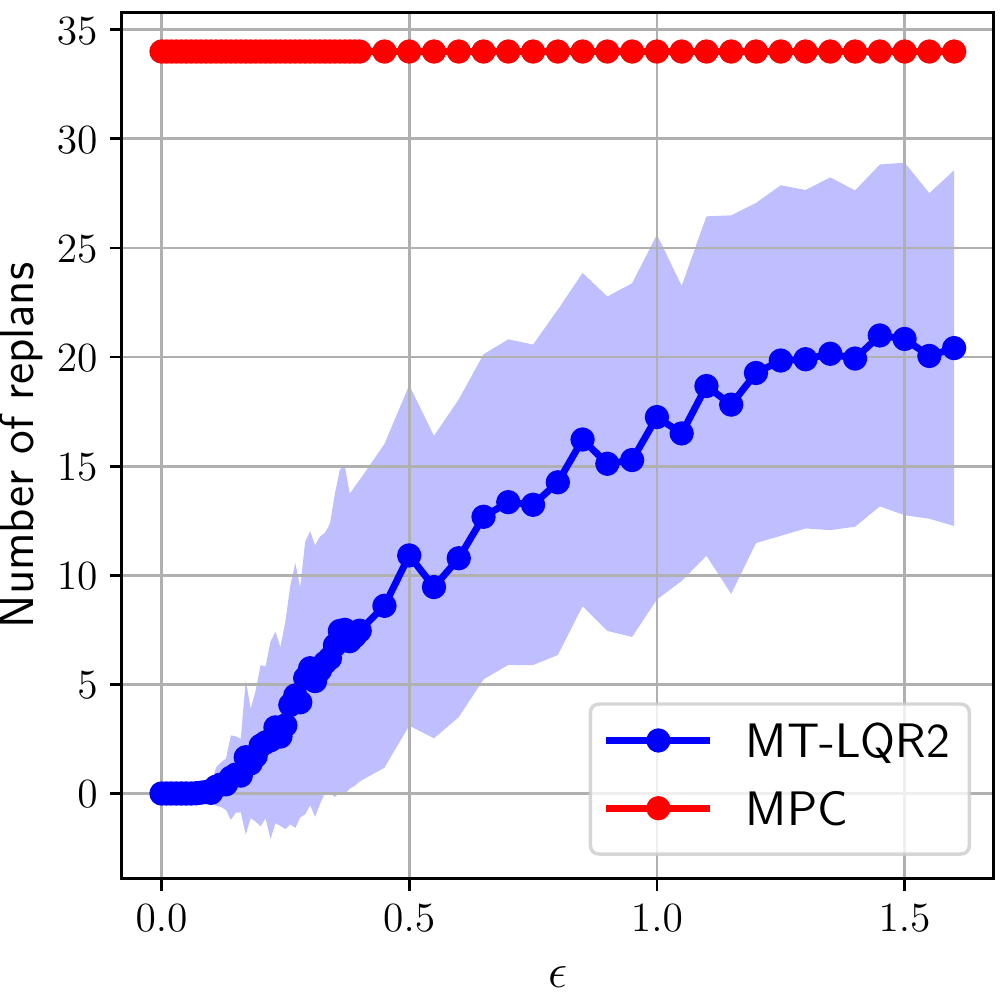}
        \caption{Three agents.}
        \label{3_agent_replan}
    \end{subfigure}
    \caption{Replanning operations vs. $\epsilon$ for $J_{\textrm{thresh}} = 2\%$}
    \label{fig:replan plot}
    \vspace*{-12pt}
\end{figure}%

\subsection{Interpretation of the results:}
From Figures~\ref{1 agent cost low} and~\ref{3 agent cost low} it can be clearly seen that the decoupled feedback law (T-LQR and MT-LQR) shows near-optimal performance compared to MPC at low noise levels~($\epsilon \ll 1$). At medium noise levels, replanning (T-LQR2 and MT-LQR2) helps to constrain the cost from deviating away from the optimal. Figure~\ref{fig:replan plot} shows the significant difference in the number of replans, which determines the computational effort, taken by the decoupled approach compared to MPC. Note that the performance of the decoupled feedback law approaches MPC as we decrease the value of $J_{\textrm{thresh}}$. The significant difference in computational time between MPC and T-LQR2 can be seen from Figure~\ref{timevsHc_1_1} which shows results for $\epsilon=0.1$. For $H_c = 35$ (i.e. we plan for the entire time horizon), and $J_{\textrm{thresh}}$= .2\% there is not much difference in the cost between them in~\ref{costvsHc_1_1} (both are in the dark green region), while there is a significant change in computation time as seen in~\ref{timevsHc_1_1}. The trend is similar in the multi-agent case as seen in Figures~\ref{fig:costvsHc_3_1} and ~\ref{fig:timevsHc_3_1} which again shows that the decoupling feedback policy is able to give computationally efficient solutions which are near-optimal in low noise cases by avoiding frequent replanning.

At high noise levels, Figures~\ref{1 agent cost full} and~\ref{3 agent cost full} show that T-LQR2 and MT-LQR2 are on a par with MPC. Additionally, we also claimed that planning too far ahead is not beneficial at high noise levels. It can be seen in Figure~\ref{fig:costvsHc_3_7} that the performance for MPC as well as MT-LQR2 is best at $H_c = 20$. Planning for a shorter horizon also eases the computation burden as seen in Figure~\ref{fig:timevsHc_3_7}. Though not very significant in the single agent case, we can still see that there is no difference in the performance as the horizon is decreased in Figure~\ref{costvsHc_1_7}. It can also be seen in Figure~\ref{3 agent cost full} where MPC-SH and MT-LQR2-SH both with $H_c=7$ outperform MPC with $H_c=35$ at high noise levels which again show that the effective planning horizon decreases at high noise levels.  

\section{CONCLUSIONS}
\label{section:conclusion}
In this paper, we have considered a class of stochastic motion planning problems for robotic systems over a wide range of uncertainty conditions parameterized in terms of a noise parameter $\epsilon$. We have shown extensive empirical evidence that a simple generalization of a recently developed ``decoupling principle" can lead to tractable planning without sacrificing performance for a wide range of noise levels. Future work will seek to treat the medium and high noise systems, considered here, analytically and look to establish the near-optimality of the scheme. Further, we shall consider the question of ``when and how to replan'' in a distributed fashion in the multi-agent setting, as well as relax the requirement of perfect state observation.






\printbibliography
\end{document}